\documentclass[10pt, aps, prb, twocolumn, superscriptaddress, amsmath, amsfonts, amssymb, amsthm]{revtex4-2}
\usepackage{graphicx}% Include figure files
\usepackage{dcolumn}% Align table columns on decimal point
\usepackage[unicode=true,
 bookmarks=false,
 breaklinks=true,pdfborder={0 0 1},backref=false,colorlinks=true]
 {hyperref}
\hypersetup{
 pdfcreator={},pdfproducer={LaTeX with hyperref},linkcolor=blue,anchorcolor=blue,citecolor=blue,filecolor=red,menucolor=red,urlcolor=blue,pdfstartview=FitV,pdfhighlight=/I,hypertexnames=true}
\makeatletter
\usepackage{lmodern}
\usepackage{mathptmx, newtxtext, newtxmath, xspace}
\usepackage{colortbl}
\usepackage{array}
\renewcommand{\arraystretch}{1.8} % Increase row height
\usepackage{bbm}
\usepackage{siunitx}
\usepackage{ulem}
\usepackage{comment}
\usepackage[table, dvipsnames]{xcolor}
\definecolor{colorA}{cmyk}{0,0,0,0.05}
\definecolor{colorB}{cmyk}{0.14,0.04,0,0}
\definecolor{colorC}{cmyk}{0.02,0.0799,0,0}
\definecolor{colorD}{cmyk}{0.099,0.14,0,0}
\definecolor{FMirrepColour}{gray}{0.4}

\usepackage{natbib}
\setcitestyle{numbers}

\makeatother
\begin{document}
\title{Elastic Quantum Criticality in Nematics and Altermagnets via the Elasto-Caloric Effect }
\author{Charles R. W. Steward }
\affiliation{Institute for Theory of Condensed Matter, Karlsruhe Institute of Technology,
Karlsruhe 76131, Germany}
\author{Grgur Palle }
\affiliation{Institute for Theory of Condensed Matter, Karlsruhe Institute of Technology,
Karlsruhe 76131, Germany}
\affiliation{Department of Physics, The Grainger College of Engineering, University of Illinois Urbana-Champaign, Urbana, Illinois 61801, USA}
\affiliation{Anthony J. Leggett Institute for Condensed Matter Theory, The Grainger College of Engineering, University of Illinois Urbana-Champaign, Urbana, Illinois 61801, USA}
\author{Markus Garst }
\affiliation{Institute for Theoretical Solid State Physics, Karlsruhe Institute
of Technology, Karlsruhe 76131, Germany}
\affiliation{Institute for Quantum Materials and Technologies, Karlsruhe Institute
of Technology, Karlsruhe 76131, Germany}
\author{J\"org Schmalian }
\affiliation{Institute for Theory of Condensed Matter, Karlsruhe Institute of Technology,
Karlsruhe 76131, Germany}
\affiliation{Institute for Quantum Materials and Technologies, Karlsruhe Institute
of Technology, Karlsruhe 76131, Germany}
\author{Iksu Jang }
\affiliation{Institute for Theory of Condensed Matter, Karlsruhe Institute of Technology,
Karlsruhe 76131, Germany}
\date{\today }
\begin{abstract}
The coupling between electronic nematic degrees of freedom and acoustic phonons is known to significantly alter the universality class of a nematic quantum critical point (QCP). While non-Fermi-liquid behaviour emerges in the absence of lattice coupling, the inclusion of interactions with acoustic phonons results in observables such as heat capacity and single-particle scattering rate exhibiting only subleading non-analytic corrections to dominant Fermi-liquid terms. In this work, we demonstrate that the elastocaloric effect (ECE) -- the adiabatic temperature change under varying strain -- and the thermal expansion deviate from this pattern. Despite lattice coupling weakening the singularity of the ECE, it preserves a dominant temperature dependence that deviates from the prediction one would obtain from Fermi liquid theory, an effect which we will refer to as an elasto-caloric anomaly. By drawing analogies between nematic systems and field-tuned altermagnets, we further show that similar responses are expected for the ECE near altermagnetic QCPs. We classify the types of piezomagnetic couplings and analyse the regimes arising from field-tuned magnetoelastic interactions. Our findings are shown to be consistent with the scaling theory for elastic quantum criticality and they further emphasize the suitability of the ECE as a sensitive probe near QCPs.
\end{abstract}
\maketitle

\section{Introduction}
Nematic quantum criticality has been discussed as one possible driving force of non-Fermi liquid behaviour~\citep{Oganesyan2001,Metzner2003,Kee2003,Dell2007,Zacharias2009,Fradkin2010,Metlitski2010,Fernandes2014,Nie2014,Klein2020},  with anomalous power-law dependencies of the single-particle lifetime or the electronic specific heat~\cite{Grossman2021} in two-dimensional systems and marginal Fermi liquid behaviour~\cite{Varma1989} in $d=3$. The same fluctuations have  been shown to cause $s$-wave superconductivity or to enhance unconventional pairing states, such as $d$-wave pairing~\citep{Lederer2015,Metlitski2015,Lederer2017,Klein2018,Klein2019}.

A special aspect of a nematic state -- i.e., a state that breaks rotation symmetry without breaking translation invariance -- is that it naturally couples to acoustic phonons of the same symmetry. This coupling is responsible for the dramatic softening of the shear modulus in nematic materials~\citep{fernandes2010effects,Bohmer2014nematic,Goto2011quadrupole}, where the renormalized elastic constant $C_{\rm ren}(T)$ vanishes at a second order nematic phase transition due to the divergence of  the nematic susceptibility $\chi_{{\rm nem}}$. In the case where  in the absence of electronic nematicity the lattice is harmonic, one may show that~\citep{fernandes2010effects}   
\begin{equation}
C_{\rm ren}^{-1}=C^{-1}+\frac{\lambda^{2}}{C^{2}}\chi_{{\rm nem}}.\label{eq::elasticrenorm_nem}
\end{equation} 
Indeed, $C_{\rm ren}$ vanishes whenever $\chi_{{\rm nem}}$ diverges.  In Eq.~\eqref{eq::elasticrenorm_nem}, $\lambda$ is the nemato-elastic coupling constant that governs the interaction of strain (i.e.\ phonon modes) with the nematic order parameter through
\begin{equation}
    H_{\rm{nem-latt}}=\lambda \int d^{3}\boldsymbol{x}\epsilon_{\Gamma_{\phi}^{+}}\left(\boldsymbol{x}\right)\phi\left(\boldsymbol{x}\right).
    \label{eq:nemel}
\end{equation}
$\Gamma_{\phi}^{+}$ is the irreducible representation of the nematic order parameter $\phi\left(\boldsymbol{x}\right)$, where the superscript indicates that it is even under time-reversal. $\epsilon_{\Gamma_{\phi}^{+}}$ is the strain field that transforms under $\Gamma_{\phi}^{+}$.
 $C$ in Eq.~\eqref{eq::elasticrenorm_nem} is the bare elastic constant for $\lambda=0$, roughly given by the high-temperature value of the elastic constant where $\chi_{\rm nem}$ is small. In Ref.~\citep{Karahasanovic2016} it was shown that, whenever the lattice softens due to nematic fluctuations, long-range strain forces act back onto the nematic modes and cause a crossover to the mean-field-like behaviour that was discussed earlier~\citep{Levanyuk1970,Cowley1976}. 

\begin{figure}
     \centering
     \includegraphics[width=0.48\textwidth]{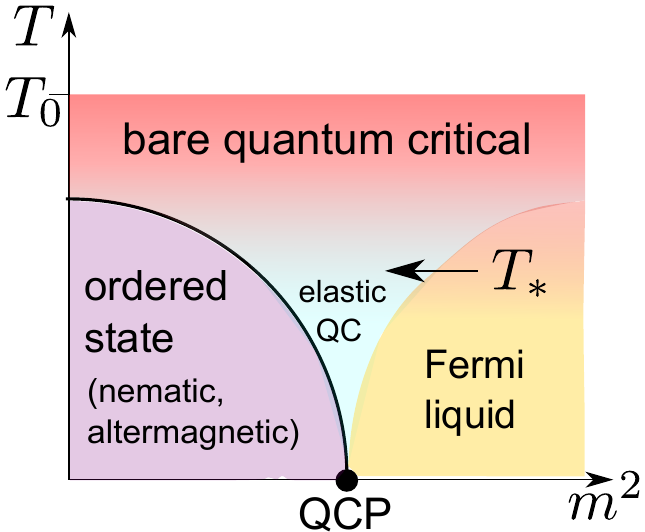}
     \caption{Phase diagram  near a nematic or altermagnetic quantum critical point (QCP), tuned by the non-thermal parameter $m^2$. The coupling to elastic degrees of freedom gives rise to a change in the universality class of the  QCP~\citep{Paul2017} from the  bare quantum critical to the  elastic quantum critical  regime at the  crossover temperature $T_{\ast}$.  For nematic QCP $T_{\ast}$ is fixed by the nemato-elastic constant, while it can
be tuned by applying an external magnetic field for altermagnets, where $T_{\ast}(H=0)=0$. The crossover to bare quantum critical behaviour occurs at $T_0$.}
     \label{fig::phase_01}
 \end{figure}

At a nematic quantum critical point (QCP) one finds, in the absence of the coupling to the lattice, deviations from Fermi liquid behaviour. For $d=3$, the Sommerfeld coefficient $\gamma(T)=c(T)/T$, with specific heat $c(T)$, diverges like $\gamma\sim \log(T_0/T)$, like in a marginal Fermi liquid~\cite{Varma1989}. The temperature scale
%\textcolor{red}{IJ:
\begin{equation}
T_0= \frac{1}{3\pi\alpha}\left(\frac{\Lambda}{k_{F}}\right)^{3}E_{F},
\label{eq:T_0}
\end{equation}
%If $E_f=\frac{1}{2}k_Fv_f$.
%}
%\begin{equation}
%T_0= \frac{1}{6\pi\alpha}\left(\frac{\Lambda}{k_{F}}\right)^{3}E_{F},
%\label{eq:T_0}
%\end{equation}
is determined by the Fermi energy $E_{\rm F}$ (more precisely the electronic energy scale below which one can safely linearize the electronic dispersion), the  bosonic cutoff $\Lambda$, the Fermi wave vector $k_F$, and the dimensionless coupling constant $\alpha$ of electrons to nematic collective modes; see Eq.~\eqref{eq:alpha} below.  

In Ref.~\citep{Paul2017} it was shown that the coupling of electronic and elastic modes also changes the universality class of a nematic quantum critical point (QCP).
Non-analytic dependencies of the excitation spectra, like the ones that occur in the absence of strain coupling, often follow from coupling to soft modes such as the condensation of a bosonic order parameter. Near such points in momentum space, the excitation energies are arbitrarily small, regardless how one approaches them. Coupling to strain leads to non-analytic behavior; modes are only soft if one approaches them along specific directions, giving rise to soft lines.  A consequence, discussed in Ref.~\citep{Paul2017}, is that the Sommerfeld coefficient saturates below the temperature
\begin{equation}
T_{\ast}\sim \frac{\lambda^{3}}{C^{3/2}J^{3}}T_{0},
\label{eq:Tstar0}
\end{equation}
if one includes the coupling to the lattice. Here, $J$ is an energy scale that characterises total bandwidth of the nematic collective mode.  $C$ is the bare elastic constant of Eq.~\eqref{eq::elasticrenorm_nem} that governs the symmetry channel of the nematic order parameter. 
Hence, below $T_\ast$ the expected divergence $\gamma\sim \log(T_0/T)$ is suppressed and replaced by $\gamma\sim \log(T_0/T_\ast)$.    Similar behaviour was shown to occur for the single-particle self energy that also changes from its marginal Fermi liquid form in the absence of the nemato-elastic coupling to Fermi-liquid-like behaviour once the lattice coupling is included~\citep{Paul2017}. The origin of this behaviour is that the sound velocity vanishes at the QCP only along isolated lines in momentum space, which gives rise to non-analytic, long-range order-parameter interactions which suppress classical and quantum fluctuations. 

Considering solely the heat capacity or the single particle lifetime, the electronic system at the QCP is not distinguishable from a Fermi liquid, at least if one concentrates on the leading low-$T$ contributions. It is therefore an interesting question to explore whether, indeed, the elastic quantum critical state behaves in every respect like a Fermi liquid or whether there are unique fingerprints of quantum criticality that are qualitatively different from the Fermi liquid expectation and that dominate at low $T$.

An interesting thermodynamic observable that has recently attracted interest because of advances in high-precision measurements~\cite{Ikeda2019ac} is the elastocaloric effect (ECE), which is quantified by
\begin{equation}
    \eta(T)=\left . \frac{dT}{d\epsilon_0} \right|_S,
    \label{eq:ECE_def}
\end{equation}
i.e., the temperature change caused by adiabatically changing the strain. Here, $\epsilon_0$ stands for strain that transforms trivially under point group operations, such as $\epsilon_{xx}+\epsilon_{yy}$ or $\epsilon_{zz}$ for a tetragonal system. The alternative scenario of an elastocaloric effect with strain $\varepsilon$ that transforms like the nematic order parameter, experimentally investigated in Ref.~\cite{ikeda2021elastocaloric}, will be discussed elsewhere~\cite{Steward2025_2}.
Measurements of $\eta(T)$ were used to reveal the entropy landscape and phase diagram of nematic systems~\cite{ikeda2021elastocaloric,Rosenberg2024nematic}, unconventional superconductors~\cite{Li2022elastocaloric,Palle2023constraints,Ghosh2024elastocaloric}, and states of complex magnetic order~\cite{Ye2023elastocaloric,ye2024measurement}.
The connection to the entropy landscape follows from the thermodynamic identity
\begin{equation}
    \eta=-\frac{\partial S\left(\epsilon_0,T\right)/\partial\epsilon_0|_T}{\partial S\left(\epsilon_0,T\right)/\partial T|_{\epsilon_0}}.
    \label{eq:ECE_deriv}
\end{equation}
The denominator is proportional to the heat capacity and hence measures energy fluctuations. The numerator can, accordingly be  interpreted as the correlation between  energy- and stress fluctuations in the system.
For a cubic system, $\epsilon_0 = \Delta V/V$ just corresponds to a volume change, and $\eta = - \Gamma/T$ is directly related to the conventional Gr\"uneisen parameter $\Gamma = \alpha/(c_V \kappa_T)$ with the specific heat $c_V$, the bulk thermal expansion $\alpha$; and the isothermal compressibility $\kappa_T$~\cite{Zhu2003}. As the Gr\"uneisen parameter is constant for a Fermi liquid, the elastocaloric effect scales linearly with temperature $\eta \sim T$ for a conventional metal. Consequently, any singular $T$-dependence of $\eta/T$ signals the presence of an elasto-caloric anomaly which is not consistent with Fermi liquid theory.

% which also reveals the close connection to the Gr\"uneisen parameter $\Gamma=\alpha/c$ with thermal expansion $\alpha$ and heat capacity $c$~\cite{Zhu2003}.

\begin{figure}
    \centering
    \includegraphics[width=\linewidth]{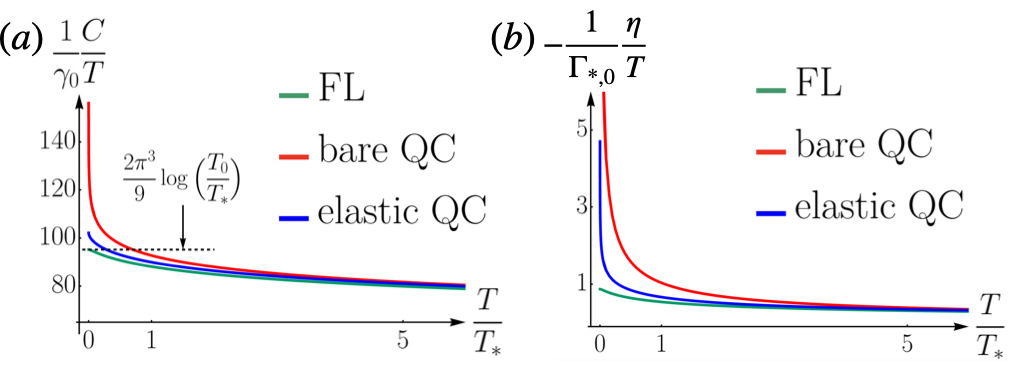}
    \caption{The heat capacity (panel (a)) and the elastocaloric coefficient (panel (b)) for the bare QCP, the Fermi liquid away from the QCP, and the elastic QCP.  For the bare QCP, the gap of the boson is zero while it is non-zero for the Fermi liquid case. For the  elastic QCP, the gap vanishes on isolated lines in momentum space only while it takes values up to  $m=m_*$ elsewhere in momentum space. For easier comparison we used a mass  $m=m_*$ for the Fermi liquid, ensuring that $T_{FL}$ (Eq.~\eqref{eq:TFL}) and $T_{*}$ (Eq.~\eqref{eq:Tstar}) are identical. The Sommerfeld coefficient  (panel (a)), diverges  as $T\rightarrow 0$ in the bare QC regime, while it remains finite for the elastic QC and the Fermi liquid.  The elastocaloric coefficient (panel (b)) is able to distinguish these two regimes, as gapless bosons which form along critical lines lead to a divergence of $\eta/T$ for both, the elastic QC and bare QC regimes. The constants $\gamma_0$ and $\Gamma_{*,0}$ are defined in Eqs.~\eqref{eq:gamma0} and~\eqref{eq:Gammastar0} respectively. The normalization by $-\Gamma_{*,0}$ in panel (b) is due to the fact that the elastocaloric coefficient and the Gr\"uneisen parameter have, by convention, opposite signs.
    }
    \label{fig::heatelasto}
\end{figure}

In this paper, we analyse the elastocaloric effect near a nematic quantum critical point. We show that, in distinction to the heat capacity or the single-particle scattering rate,  $\eta(T)$  behaves differently from the Fermi liquid expectation in the elastic quantum critical regime. Even though the temperature dependency of the elastocaloric effect changes at $T_{\ast}$, with a less singular behavior compared to the case without strain coupling,  elastic quantum criticality is strictly not a Fermi liquid state. This is surprising as the system is, from a single-particle perspective, clearly characterized by well defined quasi-particles. Non-analytic corrections to Fermi liquid behavior, that are sub-leading in the single particle response, become dominant for collective behavior such as the elasto-caloric effect. We will also show that this aspect is fully consistent with the generalized scaling behaviour of Ref.~\cite{Zacharias2015quantum} near the elastic critical regime. 
In Fig.~\ref{fig::heatelasto} and Tab.~\ref{tab:tab01} we show the heat capacity and the ECE for the Fermi liquid expectation, the bare QCP without elastic coupling, and the elastic quantum critical behaviour where this coupling has been included, respectively. For more information about numerical calculations, see the code in Ref.~\cite{Data}.

Fermi liquid theory describes fermionic quasiparticles alongside their collective (sound) modes. As shown clearly in Ref.~\cite{Paul2017}, coupling to strain re-establishes well-defined fermionic quasiparticles. One might therefore conclude that a nematic QCP—and by analogy, an altermagnetic QCP in an external field—is governed
entirely by Fermi liquid physics, with the remaining QCP-associated soft modes contributing only subleading corrections. However, our analysis reveals  that the situation is more nuanced: the elastocaloric effect exhibits behaviour that deviates from standard Fermi liquid predictions. Importantly, this deviation does not stem from an independent anomaly in the system’s elastic properties. As we demonstrate in Sec.~\ref{subsec:Elastocaloric effect}, our result can alternatively be derived from the fermionic contribution to the entropy. Taking a strain derivative, as done in the elastocaloric effect, amplifies what would otherwise appear as a subleading correction. This motivates the characterisation of such a state as a singular Fermi liquid. The behaviour below the crossover temperature  $T_{\ast}$ describes nematic quantum critical points with a spherical and with a cylindrical Fermi surface alike; the lattice dynamics is three dimensional in both cases and determines the universality class fully. The only case where one would expect  a two-dimensional non-Fermi liquid is above $T_{\ast}$ for a system with a cylindrical Fermi surface and with two-dimensional nematic correlations.

\begin{table}[h!]
\centering
\begin{tabular}{|>{\columncolor[HTML]{F5F5F5}}c|>{\columncolor[HTML]{F5F5F5}}c|>{\columncolor[HTML]{F5F5F5}}c|}
\hline
\textbf{} & \textbf{$c\left(T\right)$} & \textbf{$\eta\left(T\right)$} \\
\hline
\hline
Fermi liquid & $\quad \propto T \quad$ & $\quad \propto T \quad$ \\
\hline
bare quantum critical & $\quad \propto T\log T \quad$ & $\quad \propto T^{1/3}/\log T \quad$ \\
\hline
elastic quantum critical & $\quad \propto T \quad$ & $\quad \propto T\log T \quad$ \\
\hline
\end{tabular}
\caption{Heat capacity $c(T)$ and elastocaloric effect $\eta(T)$ as a function of temperature
in the Fermi liquid, the bare quantum critical (for $d=3$ and without coupling to the lattice), and the elastic quantum critical
regimes as indicated in Fig.~\ref{fig::phase_01}. From the perspective of the heat capacity
(and the single particle life time) the elastic QC below $T_{\ast}$
is indistinguishable from the Fermi liquid, while the elastocaloric
effect is able to discriminate between the two regimes. We depict these results in Fig.~\ref{fig::heatelasto}}\label{Table:ResultsSummary}
\label{tab:tab01}
\end{table}

\begin{figure}
     \centering
     \includegraphics[width=0.48\textwidth]{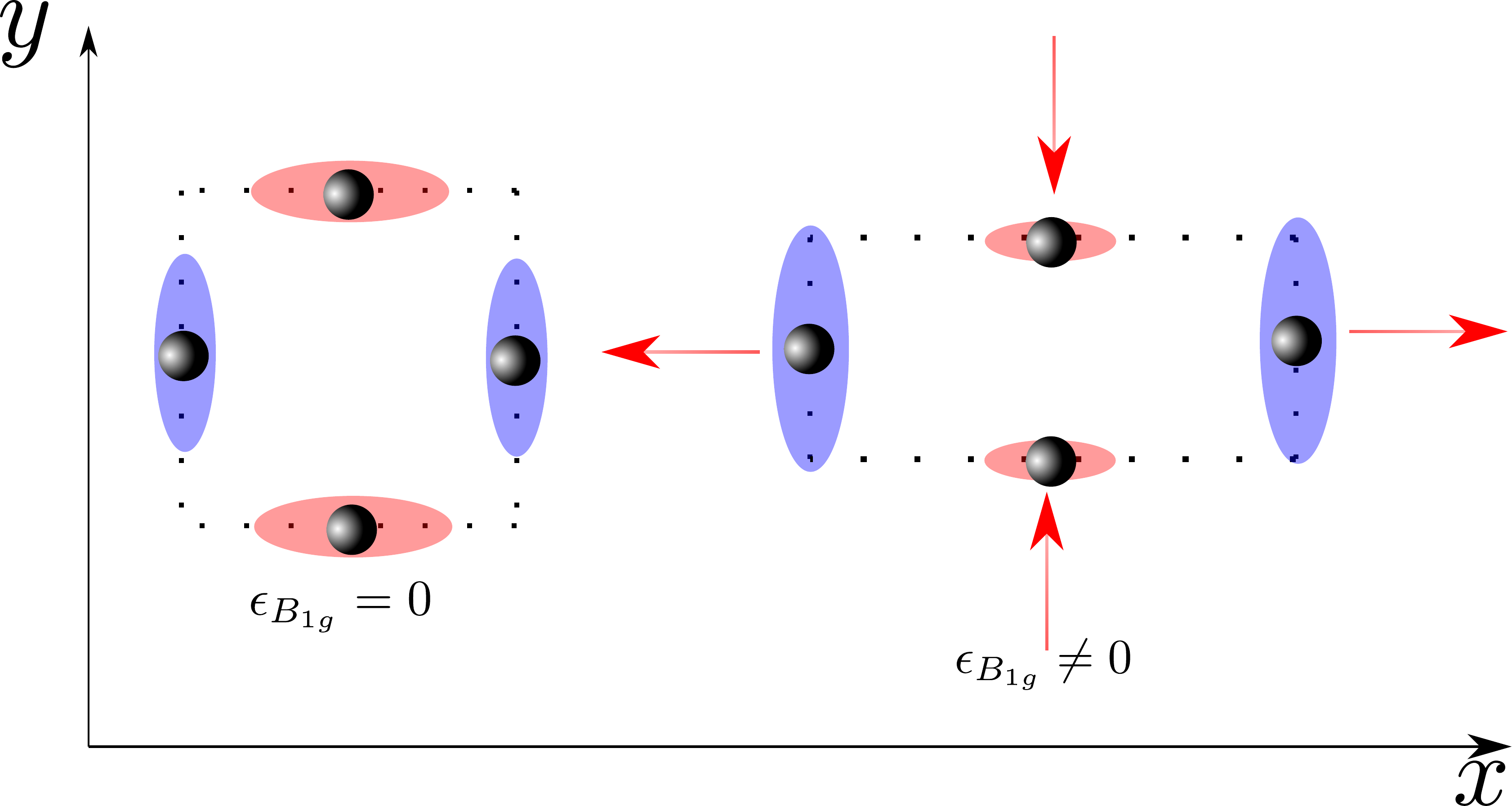}
     \caption{Illustration of the physical origin of the piezomagnetic coupling in an altermagnetic material.
     %The physical origin of the coupling considered here is simple to visualise. 
     An altermagnet which transforms according to the irreducible representation $B_{2g}$ has the order parameter $\phi_{B_{2g}}\sim \sum_{\mathbf{k}}f_{B_{1g}}\left(\mathbf{k}\right)\left\langle c^{\dag}_{\mathbf{k}}\sigma^zc_{\mathbf{k}}\right\rangle$. The form factor determines the distribution of spins within the unit cell and must transform like $B_{1g}$ given that the expectation value transforms under $A_{2g}$. The diagram in the figure hence initially appears to transform like $B_{1g}$, but is in fact $B_{2g}$. We see that by symmetry there is a zero net dipole moment within the unit cell and hence a magnetic field cannot couple. Applying a $B_{1g}$ strain breaks this symmetry, changing the spin distribution to allow for a net dipole moment in the $z$ direction such that a magnetic field aligned along the same direction can couple.}
     \label{fig::strain}
 \end{figure}

Finally, we show that  similar reasoning  can be applied to the behaviour of  altermagnetic QCPs in an external magnetic field.  Altermagnets constitute an interesting class of magnetic systems that, similar to nematics, break a point group symmetry that can be a rotation~\citep{mazin2021prediction,Smejkal2022,Smejkal2022b,Turek2022,Urru2022,Bhowal2022,vsmejkal2020crystal,mazin2023altermagnetism,feng2022anomalous,voleti2020multipolar,fiore2022modeling,betancourt2023spontaneous,winkler2023theory,yuan2021prediction,liu2022spin,bai2023efficient,vsmejkal2022anomalous,yang2021symmetry,jiang2023enumeration,mcclarty2024landau,giil2024superconductor,cheong2024altermagnetism,hodt2024spin,jaeschke2024supercell,krempasky2024altermagnetic,fernandes2024topological}. In distinction to nematic systems they also break time-reversal symmetry, yet the product of the two operations remains a symmetry.  
While in altermagnets there is no direct coupling to strain along the lines of Eq.~\eqref{eq:nemel}, in the presence of a magnetic field a piezomagnetic coupling~\citep{patri2019unveiling,aoyama2024piezomagnetic} becomes possible:
\begin{equation}
H_{\rm{am-latt}}=\sum_{i}\sum_{\alpha}\lambda^a_{\alpha,i}H_{\alpha}\int d^{3}\boldsymbol{x}\epsilon_{\Gamma_{i}^{+}}\left(\boldsymbol{x}\right)\phi^a\left(\boldsymbol{x}\right).
\label{eq::Field_coupling}
\end{equation}
It  is symmetry-allowed, provided that $\Gamma_\phi^{-}\in\Gamma_{H_{\alpha}}^{-}\otimes\Gamma_{\epsilon_i}^{+}$, i.e.\ the product representation of
the $\alpha$-th component of the magnetic field $\Gamma_{H_{\alpha}}^{-}$
and strain $\Gamma_{\epsilon_i}^{+}$ must contain the order parameter
representation $\Gamma_\phi^{-}$. A physical picture showing how a symmetry breaking strain leads to a magnetisation, which a magnetic field can couple to, is illustrated in Fig.~\ref{fig::strain} further below.
For easier readability we give a brief summary of some of the symmetry aspects of the piezomagnetic coupling in altermagnets in Sec~\ref{sec:StrainCouplingAlt}.  We also demonstrate that this coupling will change the universality class  at an altermagnetic QCP. 

As a consequence of Eq.~\eqref{eq::Field_coupling}, the coupling to elastic modes in quantum critical altermagnets, in distinction to nematic systems, can be tuned by varying an external magnetic field. This allows the crossover scale, $T_{\ast}(H)$, to be adjusted without altering the existence of the quantum critical point (QCP) itself. Below $T_{\ast}(H)$, quantum critical altermagnets are also influenced by elastic quantum criticality. In Sec.~\ref{sec::ExpAltermag}, we discuss how to detect the elastic QCP, leveraging the tunability provided by the magnetic field in altermagnetic systems.

The remainder of this paper is organized as follows.
In   section~\ref{sec:nematicQCP} we analyse elastic quantum criticality at a nematic QCP. In   section~\ref{sec:StrainCouplingAlt} we  summarize the piezomagnetic coupling in altermagnets. If one is not  concerned with quantum critical fluctuations, this section can be read as a stand-alone discussion of this coupling. Section~\ref{sec:altQCP} discusses elastic quantum criticality at an altermagnetic QCP and experimental signatures to detect it using its tunability by magnetic field. To put our results in context, in section~\ref{sec:scaling} we discuss our results for the elastocaloric effect in terms of the scaling theory of Ref.~\cite{Zacharias2015quantum}.

\section{Elastic quantum criticality at a nematic QCP}
\label{sec:nematicQCP}

In this section, we address nematic QCPs in the presence of a coupling to elastic degrees of freedom, and we extend the analysis of Ref.~\cite{Paul2017} to the elastocaloric effect $\eta$. Whereas the elastic coupling induces Fermi-liquid behaviour in the specific heat at lowest temperature at the QCP, we show, as our main finding of this section, that the elastocaloric effect still exhibits quantum critical behaviour with a singular $T$-dependence, $\eta/T \sim \log T$. To be specific about our system, we consider in this section a tetragonal crystal which transforms according to the $D_{4h}$ point group. However, our results are  applicable to other systems with different point groups, as long as they describe one-dimensional irreducible representations. In Sections \ref{subsec:DirectionSelectiveCriticality}, \ref{sec::nem_heat}, and \ref{subsec:FermionicSelfEnergy} we first review results previously obtained in the literature. Section \ref{subsec:Elastocaloric effect} presents our findings on the elastocaloric effect.

\subsection{Direction selective criticality}
\label{subsec:DirectionSelectiveCriticality}

% We first investigate how the elastic coupling affects  quantum criticality in nematic systems, extending the analysis of Ref.~\cite{Paul2017} to the elastocaloric effect. To be specific about our system, we consider a tetragonal crystal which transforms according to the $D_{4h}$ point group. However, our results are  applicable to other systems with different point groups, as long as they describe one-dimensional irreducible representations.  

% To investigate the effect of quantum fluctuations, we consider the one-loop corrections to the boson and fermion.

Following~\citep{Karahasanovic2016,Paul2017}, we consider the following action describing a nematic QCP and its coupling to elastic degrees of freedom of a tetragonal crystal,
\begin{equation}
\mathcal{S}=\mathcal{S}_c+\mathcal{S}_{\phi}+\mathcal{S}_\epsilon+\mathcal{S}_{\epsilon-\phi}+\mathcal{S}_{c-\phi}.\label{eq::LEES}
\end{equation}
The part describing free fermions 
%which consists of the part that describes free fermions 
with band dispersion $\xi(\mathbf{k})$ is given by
\begin{equation}
   \mathcal{S}_{c}=-\sum_{\sigma=\uparrow,\downarrow} \int_k c_\sigma^\dagger(k)(i\omega-\xi(\mathbf{k}))c_\sigma(k), 
\end{equation}
with $\int_k\cdots=T  \sum_{\omega} \int\frac{d^3k}{(2\pi)^3}\cdots$ and $k=(\omega,\mathbf{k})$ combines frequency and momenta. With the exception of Landau damping, that we briefly discuss below, our results will not depend  on the details of the dispersion $\xi(\mathbf{k})$.
We consider a spherical Fermi surface  in the main text. Our results are unchanged if we consider a cylindrical Fermi surface because the critical lines of soft phonons are in the 
 %$x-y$ 
 crystallographic $a-b$ plane where the symmetries of the two Fermi surfaces are identical. Hence, for now we  use $\xi(\mathbf{k})=\frac{|\mathbf{k}|^2}{2m}-\mu$. We discuss systems with cylindrical Fermi surfaces in Appendix~\ref{App:CylindricalFS}.
The part of the action that describes the nematic Ising order parameter is
\begin{equation}
    \mathcal{S}_{\phi}=\frac{1}{2} \int_q \phi(q)\Big(m^2+\omega^2+v_\phi^2|\mathbf{q}|^2\Big)\phi(-q),
\end{equation}
where $v_\phi$, which has dimension of a velocity, determines the spatial stiffness of nematic fluctuation. This should be supplemented by the $\phi^4$ interaction of the Ising degree of freedom. As this interaction does not lead to important contributions, we omit it in the following discussion.
%We do not include non-linear interactions $\sim\phi^4$ of the Ising degree of freedom as it is an irrelevant correction at the QCP. 
The elastic degrees of freedom are determined by
\begin{equation}
    \mathcal{S}_{\epsilon}=\int_x\Big[\frac{\rho}{2}\sum_{i=x,y,z}(\partial_\tau u_i)^2+F_{\rm el}[\{u_i\}]\Big],
\end{equation}
where $\rho$ is the mass density, $\int_x\cdots =\int d\tau \int d^3x \cdots$, and the elastic energy 
\begin{align}
F_{\rm el}[{u_{i}}]&=\frac{C_{11}}{2}(\epsilon_{xx}^2+\epsilon_{yy}^2)+\frac{C_{33}}{2}\epsilon_{zz}^2+2C_{44}(\epsilon_{xz}^2+\epsilon_{yz}^2)\nonumber\\
&+2C_{66}\epsilon_{xy}^2+C_{12}\epsilon_{xx}\epsilon_{yy}+C_{13}(\epsilon_{xx}+\epsilon_{yy})\epsilon_{zz},
\end{align}
is determined by the usual elastic constants, $C_{ij}$, given in Voigt notation and with strain field $\epsilon_{ij}=\tfrac{1}{2}\left(\partial_i u_j+\partial_j u_i\right)$ expressed in terms of the displacement $u_i$. This term determines the dispersion relations and polarisation vectors for the elastic modes. For example, with the assumptions $C_{13}=-4C_{44}$ and $C_{66}=\frac{C_{11}-C_{12}}{8}$~\citep{Steward2023}, we obtain the  elastic modes $\Omega_i(\mathbf{q})$:
\begin{eqnarray}
\rho \Omega^2_1(\mathbf{q})&=&4C_{44}|\mathbf{q}_{2d}|^2+C_{33}q_z^2, \nonumber \\
\rho \Omega^2_2(\mathbf{q})&=& C_{11}|\mathbf{q}_{2d}|^2+4C_{44}q_z^2  , \nonumber \\
\rho \Omega^2_3(\mathbf{q})&=&  \frac{C_{11}-C_{12}}{2}|\mathbf{q}_{2d}|^2+4C_{44}q_z^2,
\label{eq::SimplifiedElasticModes}
\end{eqnarray}
where the polarization vectors of the  elastic modes are $\vec{\lambda}_1(\mathbf{q})=(0,0,1)^T$, $\vec{\lambda}_2(\mathbf{q})=\frac{1}{|\mathbf{q}_{2d}|}(q_x,q_y,0)^T$, and $\vec{\lambda}_3(\mathbf{q})=\frac{1}{|\mathbf{q}_{2d}|}(-q_y,q_x,0)^T$, respectively, with $\mathbf{q}_{2d}=\left(q_x,q_y\right)$.

For a nematic order parameter that transforms according to $B_{1g}$, the strain coupling of Eq.~\eqref{eq:nemel} takes the form
\begin{equation}
 \mathcal{S}^{}_{\epsilon-\phi}=\frac{\lambda}{2} \int_x \phi(x)\left(\epsilon_{xx}(x)-\epsilon_{yy}(x)\right). \label{eq::Action}  
\end{equation}
The nemato-elastic coupling strength  $\lambda$ has  units of $\text{energy}^{3/2}\times\text{length}^{-3/2}$. The direct coupling of the order parameter to electrons is 
\begin{equation}
    \mathcal{S}_{c-\phi}=g\int_{k,q}h_{\boldsymbol{k}}c_{k+\frac{q}{2}\sigma}^{\dagger}c_{k-\frac{q}{2}\sigma}\phi_{q},
\end{equation}
where $g$ is the coupling constant of the problem with units  $\text{energy}^{3/2}\times\text{length}^{3/2}$. The form factor $h_{\boldsymbol{k}}$ transforms like the order parameter under rotations, while its magnitude is set such that $\left\langle h_{\boldsymbol{k}}^{2}\right\rangle _{{\rm FS}}=1$ if averaged over the Fermi surface.
$S_{c-\phi}$  determines the nematic dynamics through Landau damping.

Ignoring for  the moment the nemato-elastic coupling, the one-loop analysis of the remaining problem is standard and yields for $d=3$ the marginal Fermi liquid behaviour for the electronic self energy~\citep{Oganesyan2001,Metzner2003,Kee2003,Dell2007,Zacharias2009,Fradkin2010,Metlitski2010,Fernandes2014,Nie2014,Klein2020} 
\begin{equation}
    \Sigma\left(\omega\right)=-\alpha\left(\omega\log\left(\frac{T_{0}}{\left|\omega\right|}\right)-i\frac{\pi}{2}\left|\omega\right|\right),
\end{equation}
and  the Sommerfeld coefficient
\begin{equation}
    \gamma=\alpha\frac{2\pi^{2}}{3}\rho_{F}\log\frac{T_{0}}{T}.
    \label{eq:gamma_0}
\end{equation}
where a factor of $2$ due to spin is included. The temperature scale $T_0$ is given in Eq.~\eqref{eq:T_0}. The  dimensionless coupling constant is given by
\begin{equation}
    \alpha=\frac{g^{2}}{12\pi^{2}v_{F}v_{\phi}^{2}}.
    \label{eq:alpha}
\end{equation}
Finally,  the strain dependence of the entropy reads
\begin{equation}
    \frac{\partial S}{\partial\epsilon_{0}}=-b\Lambda^3 \frac{\frac{\partial m^{2}}{\partial\epsilon_0}}{v_{\phi}^{2}\Lambda^{2}}\left(\frac{T}{T_{0}}\right)^{1/3}
    \label{eq:dSde_bare}
\end{equation}
with numerical coefficient $b=\frac{2\Gamma\left(\frac{4}{3}\right)\zeta\left(\frac{4}{3}\right)}{3^{5/2}\pi^{2}}\approx 0.042$. This is fully consistent with the findings of Ref.~\cite{Zhu2003}. Here we used that the dominant dependence on symmetric strain is the change of  $m^2$ that tunes the distance to the QCP. Using $\eta=\gamma^{-1}\frac{\partial S}{\partial\epsilon_{0}}$ for the elasto-caloric effect yields the $T^{1/3}/\log T$ behaviour listed in Tab.~\ref{tab:tab01}. The overall coefficient $\Lambda^3 $ in Eq.~\eqref{eq:dSde_bare} merely accounts for the fact that we are analyzing the entropy density and the characteristic length scale is the inverse boson cutoff $\sim 1/\Lambda$. Furthermore, the  coefficient $\frac{\partial m^{2}}{\partial\epsilon_0}/(v_\phi^2 \Lambda^2)$ is a dimensionless measure of the overall strain sensitivity of the system and will be encountered frequently in the subsequent analysis.

%Before calculating the one-loop correction from the lattice-boson coupling ($S_{\epsilon-\phi}$), 
%\begin{subequations}
%\begin{align}
%&\lambda_1(\mathbf{q})=c_{44}|\mathbf{q}_{2d}|^2+c_{11}q_z^2,\; \vec{\lambda}_1(\mathbf{q})=(0,0,1)^T,\\
%&\lambda_2(\mathbf{q})=c_{11}|\mathbf{q}_{2d}|^2+c_{44}q_z^2,\; \vec{\lambda}_2(\mathbf{q})=\frac{1}{|\mathbf{q}_{2d}|}(q_x,q_y,0)^T,\\
%&\lambda_3(\mathbf{q})=\frac{c_{11}-c_{12}}{2}|\mathbf{q}_{2d}|^2+c_{44}q_z^2,\; \vec{\lambda}_3(\mathbf{q})=\frac{1}{|\mathbf{q}_{2d}|}(-q_y,q_x,0)^T,\label{eq::thirdElasticMode}
%\end{align}
%\label{eq::SimplifiedElasticModes2}
%\end{subequations}
%where $\lambda_i(\mathbf{q})$ and $\vec{\lambda}_i(\mathbf{q})$ are the dispersion relation and the polarisation vector of the $i$th elastic mode respectively and $\mathbf{q}_{2d}=\left(q_x,q_y\right)$. 

\begin{figure}
\centering
\includegraphics[scale=0.3]{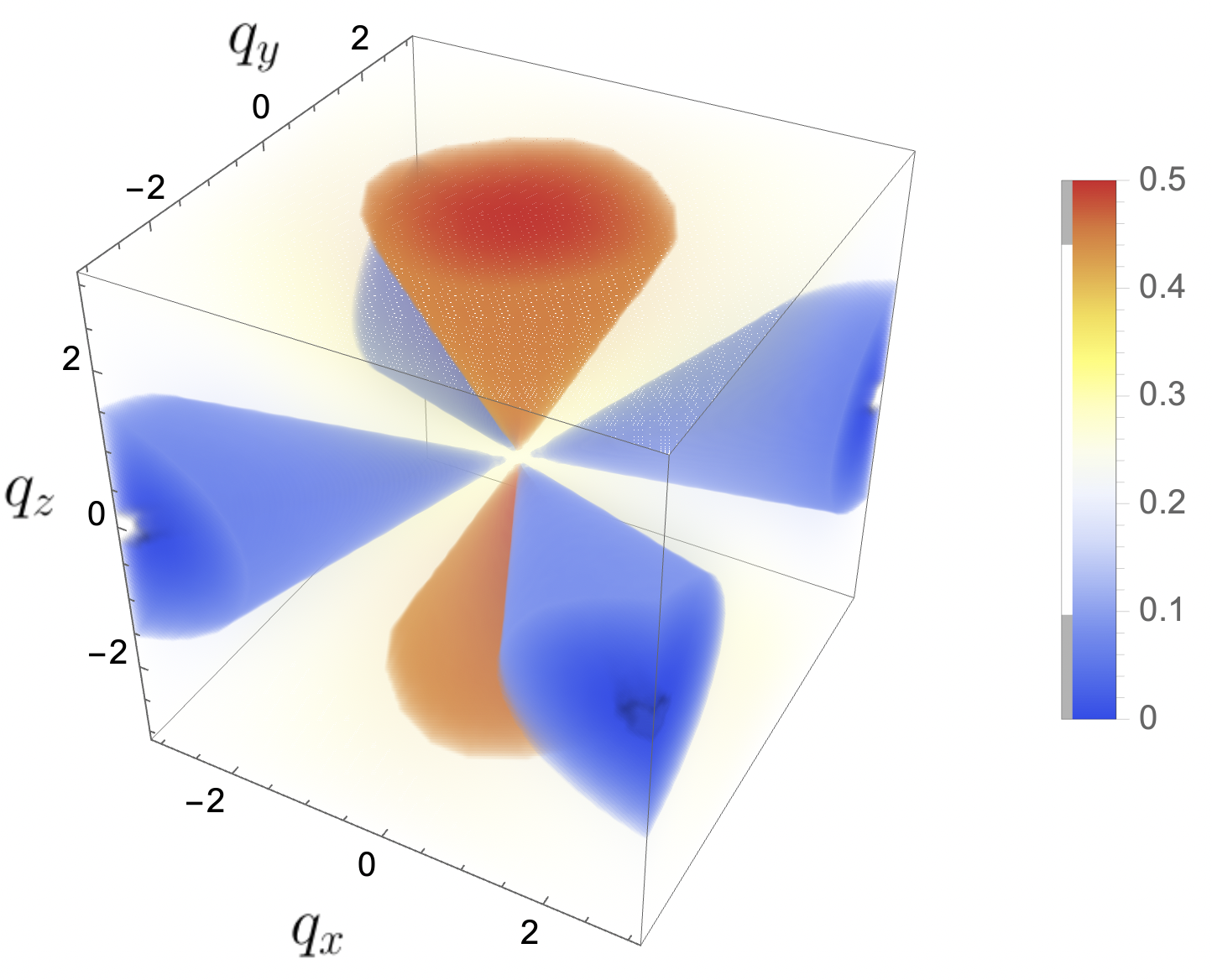}
\caption{Density plot of $\frac{\rho}{\lambda^2} \delta m^2(q_x,q_y,q_z)$ with $C_{11}=1$, $C_{12}=0$ and $C_{44}=1/4$. Soft lines in momentum space  and their vicinity are indicated in blue, while red marks regions with a particularly hard gap.}\label{fig:DirectionSelectiveMass}
\end{figure}

Next, we analyse how these results change as one includes the nemato-elastic coupling of Eq.~\eqref{eq::Action}. We can integrate out exactly the elastic modes and  obtain the following additional term in the order-parameter action:
\begin{eqnarray}
\Delta \mathcal{S}_{\phi}&=&-\frac{\lambda^2}{8 \rho}\int_q \phi(q)\phi(-q)\nonumber\\
& \times & \Bigg[\frac{1}{|\mathbf{q}_{2d}|^2}\frac{\left(q_x^2-q_y^2\right)^2}{\omega^2+ \Omega^2_2(\mathbf{q})}+\frac{1}{|\mathbf{q}_{2d}|^2}\frac{4 q_x^2q_y^2}{\omega^2+ \Omega^2_3(\mathbf{q})}\Bigg].\label{eq::DeltaSphiFromelasticity}
\end{eqnarray}
As discussed in Refs.~\citep{Paul2017,Karahasanovic2016,Steward2023}, after a small-$\omega$ expansion, this correction renormalizes the mass, which now becomes a function of the direction of the momentum,  and hence results in direction-selective criticality~\cite{Paul2017}.
The direction-dependent mass is given by
\begin{align}
m^2(\mathbf{q})&=m^2- \frac{\lambda^2}{4\rho}\frac{1}{|\mathbf{q}_{2d}|^2}\left(\frac{(q_x^2-q_y^2)^2}{\Omega^2_2(\mathbf{q})}+\frac{4q_x^2q_y^2}{\Omega^2_3(\mathbf{q})}\right).\label{eq::RenormalizedMassnem}
\end{align}
Using Eq.~\eqref{eq::SimplifiedElasticModes}, we find that the minimum value of the mass is given along  $q_x=\pm q_y$ in the $q_z=0$ plane with the following value:
\begin{align}
\left. m^2(\mathbf{q})\right|_{\rm{min}}=m^2- \frac{\lambda^2}{2}\frac{1}{C_{11}-C_{12}},\label{eq::rminHzNeqZero}
\end{align}
which  shifts the critical point from $m=0$ to $m_{\rm{min}}=0$~\cite{Paul2017}, i.e.\ it sets the scale 
\begin{equation}
    m_\ast^2=\frac{\lambda^2}{2}\frac{1}{C_{11}-C_{12}}.
    \label{eq:mast}
\end{equation}
Fig.~\ref{fig:DirectionSelectiveMass} shows the anisotropy in momentum space of the renormalized mass at the critical point. It  shows two critical lines: $q_x=q_y$ and $q_x=-q_y$ in the $q_z=0$ plane. Along the hard directions, like $q_x=q_y=0$, the renormalized mass is at least $m_\ast$.   For our subsequent analysis of the free energy we assume the system is positioned at the elastic quantum critical point ($m=m_\ast$) and approximate
the $m(\mathbf{q})$ of Eq.~\eqref{eq::RenormalizedMassnem}. We expand  the momentum dependence of $m(\mathbf{q})$ close to the critical lines and far away from them, and we treat these distinct contributions separately:
\begin{align}
    m^2(\mathbf{q})\approx m_\ast^2 \left\{ \begin{array}{ll}\frac{ q_2^2+ q_z^2}{q_1^2}, & \left(\it{i}\right)\, |q_1|\gg |q_2|,|q_z|\\
   \frac{ q_1^2+ q_z^2}{q_2^2}, & \left(\it{ii}\right)\, |q_2|\gg |q_1|,|q_z|\\
    1, & \left(\it{iii}\right)\, |q_z|\gg |q_1|,|q_2|
    \end{array}\right. .
    \label{eq:rapprox}
\end{align}
Here, $q_1=\frac{q_x+q_y}{\sqrt{2}}$ and $q_2=\frac{q_x-q_y}{\sqrt{2}}$. Regimes $\left(\it{i}\right)$ and $\left(\it{ii}\right)$ describe the soft lines, while the hard, gapped part of the spectrum corresponds to $\left(\it{iii}\right)$. To gain a better intuition about what is meant by a soft line, consider as an example the static propagator at $m=0$ of the form given in Ref.~\cite{Karahasanovic2016}
\begin{equation}
\chi_{{\rm elast}}\left(\boldsymbol{q},0\right)\propto\frac{1}{q^{2}+c_{1}\frac{\left(q_{x}^{2}-q_{y}^{2}\right)^{2}}{q^{4}}+c_{2}\frac{q_{z}^{2}}{q^{2}}}.
\end{equation}
Even at the critical point this susceptibility only diverges for $\boldsymbol{q}\rightarrow\boldsymbol{0}$
if one approaches this point along the direction $\boldsymbol{q}\parallel\left(1,1,0\right)$ or the symmetry equivalent direction $\boldsymbol{q}\parallel\left(1,-1,0\right)$.
Approaching the same point, $\boldsymbol{q=0}$, from any other direction
yields a finite response. Approaching from the direction $\boldsymbol{q}\parallel\left(1,1,1\right)$
yields $\chi_{{\rm elast}}\left(\boldsymbol{q}\rightarrow\boldsymbol{0},0\right)=\frac{3}{c_{2}}$
while $\chi_{{\rm elast}}\left(\boldsymbol{q}\rightarrow\boldsymbol{0},0\right)=\frac{1}{c_{1}}$
if the momentum approaches zero along $\boldsymbol{q}\parallel\left(1,0,0\right)$.
Hence there is only one line in momentum space along which an actual
divergence of the susceptibility occurs.

In what follows, we will analyse the heat capacity as well as the elastocaloric effect by taking this anisotropic mass due to the nemato-elastic coupling into account.

We briefly discuss the renormalization of the boson dynamics as a result of coupling to fermions. As discussed in Ref.~\cite{Paul2017}, it depends on the geometry of the Fermi surface whether one can kinematically realize Landau damping. For example in single-band systems, appropriate to the cuprates, no Landau damping occurs at leading order in perturbation theory while in multiband models, appropriate to e.g.\ the iron-based materials, this is the case. In the latter case  it leads to the order-parameter action
\begin{equation}
    \mathcal{S}_{\phi}=\frac{1}{2} \int_q \phi(q)\chi(q)^{-1}\phi(-q),
\end{equation}
with  susceptibility 
\begin{equation}
   \chi\left(\mathbf{q},\omega\right)= \Big(m^2(\mathbf{q})+\omega^2+v_\phi^2|\mathbf{q}|^2+\Xi^2 \frac{|\omega|}{v_{\rm F}q}\Big)^{-1}.
   \label{eq::OneLoopBosonDispersion}
\end{equation}
where $\Xi=\sqrt{6\pi\alpha} v_{\phi}k_{F}$ is an energy scale that determines the strength of Landau damping.  In what follows, we will assume that the system is governed by Landau damping.

In our analysis of elastic modes, we used simplifying assumptions for the elastic constants. As a consequence, the  elastic mode $\Omega_3(\mathbf{q})$ of Eq.~\eqref{eq::SimplifiedElasticModes} leads to a soft plane at $q_z=0$ when $C_{11,{\rm ren}}\approx C_{12,{\rm ren}}$. This is a fine-tuned result. In the generic case only the lines with $q_x=\pm q_y$, $q_z=0$ form the softening manifold~\citep{Paul2017,Karahasanovic2016} for the $D_{4h}$ symmetry group. Despite the simplifying assumptions made, we show in Appendix~\ref{App:RenormalizedMassGeneral} that it yields the same qualitative behaviour compared to the case where we use generic expressions for the elastic modes.  This is due to the form factor $\vec{\lambda}_3(\mathbf{q})$, involved in the renormalized mass of the boson, which lifts the degeneracy to the lines $q_x=\pm q_y$, $q_z=0$.

\subsection{Specific heat at a nematic QCP}\label{sec::nem_heat}
We first analyse the  specific heat at low temperatures. Previously, this analysis was performed  in Ref.~\cite{Paul2017} and so we include a short summary here with details of the calculation given in Appendix.~\ref{app::specific_heat}. It is included here to simplify the comparison with the elastocaloric effect that we discuss next. 
%We start from the order-parameter susceptibility 
%\begin{align}
%\chi^{-1}(\omega,\mathbf{q})&\approx v_\phi^2|\mathbf{q}|^2+r(\mathbf{q})+\zeta \frac{|\omega|}{v_{\rm{F}}|\mathbf{q}|},
%\end{align}
%As a result,
%\begin{align}
%    \chi^{-1}(\Omega,\mathbf{q})\propto \frac{|\mathbf{q}|^2}{k_{\rm{F}}^2}+\frac{r(\mathbf{q})}{\zeta}+\gamma \frac{|\Omega|}{v_{\rm{F}}|\mathbf{q}|}.\label{eq::OneLoopBosonDispersion}
%\end{align}
The free energy density of the system is given by
\begin{align}
    &F=\frac{T}{2}\sum_{\omega}\int\frac{d^3q}{(2\pi)^3}\log\chi^{-1}(\omega,\mathbf{q})\nonumber \\
    &= -\int\frac{d^3q}{(2\pi)^3}\int_0^\infty \frac{d\omega }{2\pi}\coth\frac{\beta \omega}{2} \tan^{-1}\frac{\tfrac{\Xi^2 \omega}{|\mathbf{q}|v_{\rm{F}}}}{v_{\phi}^2|\mathbf{q}|^2+m^2(\mathbf{q})},
    \label{eq:free_eng}
\end{align}
where the Matsubara frequency summation is evaluated via a contour integral. 
From this we calculate the Sommerfeld coefficient $\gamma(T)=-\frac{\partial F^2}{\partial^2T}$.

We introduce 
\begin{equation}
    \gamma_0=\frac{3}{\pi}\alpha\rho_{F}. 
    \label{eq:gamma0}
\end{equation}
% \begin{eqnarray}
% \gamma & = & \gamma_0g(\bar{T},\bar{\Lambda})
%    \label{eq:gamma_nem1}
% \end{eqnarray}
% where 
% \begin{gather}
% \gamma_0=\frac{3}{\pi}\alpha\rho_{F},\label{eq:gamma0}\\
% g(\bar{T},\bar{\Lambda})=\int_{0}^{\infty}\frac{xdx}{e^{x}-1} \int_{0}^{\bar{\Lambda}}d^3\bar{q}\frac{|\bar{q}|^3\left(\bar{m}^{2}\left(\boldsymbol{q}\right)+\bar{q}^{2}\right)^{3}}{\left(\bar{q}^{2}\left(\bar{m}^{2}\left(\boldsymbol{q}\right)+\bar{q}^{2}\right)^{2}+x^{2}\bar{T}^2\right)^{2}}
%    \label{eq:gamma_nem}
% \end{gather}
% and 
and \color{black}the following dimensionless variables
\begin{gather}
\bar{m}\left(\boldsymbol{q}\right)=\frac{m\left(\boldsymbol{q}\right)}{m_{\ast}},\quad\boldsymbol{\bar{q}}=\frac{v_{\phi}\boldsymbol{q}}{m_{\ast}},
    \quad \bar{T}=\frac{T}{T_{\ast}}
    \label{eq:dimless_para}
\end{gather}
as well as $\bar{\Lambda}=\left(T_{0}/T_{\ast}\right)^{1/3}$, and the crossover temperature scale
\begin{equation}
 T_{\ast}=\left(\frac{m_{\ast}}{v_{\phi}\Lambda}\right)^{3}T_{0},
\label{eq:Tstar}
\end{equation}
 where the third power of the coefficient  is due to  the dynamical scaling exponent of the problem being $z=3$. If we use $m_\ast$ of Eq.~\eqref{eq:mast} and $J=v_{\phi}\Lambda$ with typical bandwidth $J$ of the collective nematic mode, we recover the expression for  $T_\ast$,  given in Eq.~\eqref{eq:Tstar0}. $T_\ast$ vanishes in the limit of zero nemato-elastic coupling $\lambda$. In this limit, the analysis of the integral  is straightforward  and gives the logarithmic dependence of the Sommerfeld coefficient of Eq.~\eqref{eq:gamma_0}. In this paper we always assume that $T_\ast<T_0$, i.e.\ the elastic couplings are sufficiently weak. For $T\ll T^{\ast}$, the integral simplifies to
\begin{align}
\gamma &\approx \gamma_0 \frac{\pi^2}{6} \int_{0}^{\bar{\Lambda}}d^3\bar{q}
\frac{1}{\bar q (\bar{m}^{2}\left(\boldsymbol{q}\right)+\bar{q}^{2})}.
\end{align}
which is convergent, even along the soft directions. Once again, for the full analysis please see Appendix \ref{app::specific_heat}\color{black}

In summary, for the specific heat coefficient we thus obtain
\begin{align}
    \gamma(T) \approx \frac{2\pi^{3}}{9}\gamma_0 \begin{cases}
    \log \frac{T_0}{T_\ast}, & T\ll T_{\ast},\\
    \log \frac{T_0}{T}, & T\gg T_{\ast}.
    \end{cases}
    \label{eq:Sommerfeld_el}
\end{align}
Below the crossover temperature $T_\ast $ the specific heat is Fermi liquid-like, while above $T_\ast $  the system is in the bare QC regime and behaves like a marginal Fermi liquid,  as reported in Ref.~\cite{Paul2017}.

\subsection{Fermionic self energy at a nematic QCP}
\label{subsec:FermionicSelfEnergy}
In order to investigate the behaviour around the critical point, another quantity considered in Ref.~\cite{Paul2017} is the self-energy of the electrons due to interactions with critical nematic bosons. The self-energy is given by
\begin{equation}
\Sigma\left(\mathbf{k},\omega_n\right)\propto h_{k}^2\int_{q,\Omega_n}G_{k+q}\left(\omega_n+\Omega_n\right)\chi_{q}\left(\Omega_n\right),
\end{equation}
where $G_k\left(\omega_n\right)$ is the electron propagator, $\chi_q\left(\omega_n\right)$ is the boson propagator, $h_k$ is the form factor discussed in the Landau damping section and  $k\gg q$. For the iron-based superconductor Fermi-surface, Landau damping is present. Far away from the critical directions ($q_z\gg q_{1,2}$), the susceptibility is given by Eq.~\eqref{eq::OneLoopBosonDispersion}  with $m(\mathbf{q}) \approx m$. 
Calculating the self-energy again reveals the same crossover scale $T_{\ast}$ of Eq.~\eqref{eq:Tstar} above which $m^2$ can be neglected and the self-energy acquires a non-Fermi liquid form~\citep{Paul2017,garst2010electron,metzner2003soft} $\Sigma\left(i\omega_n\right)\propto \left|\omega\right|^{2/3}$ in 2D and $\Sigma\left(i\omega_n\right)\propto \omega_n\log\left|\omega_n\right|$ in 3D. Below this temperature the corrections are Fermi liquid-like, in agreement with the calculation of the specific heat.

At low energies we need to take the contributions from the critical directions into account. We would expect hot spots in the regions where the critical directions are parallel to the Fermi surface, and we indeed find that there exists a non-analytic, subleading contribution to the self energy $\Sigma\left(i\omega\right)\propto i\omega\left(\left|\omega\right|/T_{\ast}\right)^{1/3}$. These hot spots survive due to the shape of the Fermi surface, in combination with the structure of the form factor~\cite{Paul2017}. We see that this analysis agrees with the results of the specific heat, i.e., there exists a scale $T_{\ast}$ below which the system behaves like a Fermi liquid. 
%{\color{red} THIS SHOULD READ "FERMI LIQUID" I GUESS?}

\subsection{Elastocaloric effect at a nematic QCP}
\label{subsec:Elastocaloric effect}
Having discussed the results of Ref.~\cite{Paul2017}, we see that it would appear that the coupling of a nematic order parameter to strain leads to an area of Fermi liquid behaviour around the QCP. However, we demonstrate next that the elastocaloric effect of Eq.~\eqref{eq:ECE_def} deviates from Fermi-liquid expectation and shows an anomalous behaviour even below the crossover temperature $T_*$. This anomalous behaviour of $\eta$ at lowest $T$ is, in fact, characteristic of elastic quantum criticality.

Using Eq.~\eqref{eq:ECE_deriv}, the elastocaloric effect can be expressed as follows:
\begin{align}
    \eta=-\frac{T}{c}\frac{\partial S}{\partial \epsilon}=-\frac{1}{\gamma}\frac{\partial S}{\partial \epsilon},
\end{align}
where $c$ and $\gamma$ are, respectively, the specific heat at constant strain and the Sommerfeld coefficient, which were calculated in the previous subsection.
The free energy density $F(T,\epsilon)$ is given by Eq.~\eqref{eq:free_eng} with $m(\mathbf{q})\rightarrow m(\mathbf{q},\epsilon)$, 
%\begin{align}
%    F(T,\epsilon)\propto \frac{T}{2}\sum_\Omega \int\frac{d^3q}{(2\pi)^3}\log\Big[\frac{|\mathbf{q}|^2}{k_{\rm{F}}^2}+\frac{r(\mathbf{q},\epsilon)}{\zeta}+\frac{|\Omega|}{v_{\rm{F}}|\mathbf{q}|}\Big],
%\end{align}
where $m(\mathbf{q},\epsilon)$ is the renormalized mass which depends on the strain $\epsilon$  that transforms trivially under point group operations, such that $ \left.\partial m^2\left(\mathbf{q},\epsilon\right)/\partial\epsilon\right|_{\epsilon=0}\neq0 $. Calculating the necessary derivatives, we find
\begin{comment}
\begin{align}
    &-\frac{\partial S}{\partial \epsilon}=-\frac{\partial F^2}{\partial \epsilon\partial T}\equiv \frac{\Lambda^3}{32\pi^4 m_{\ast}^{2}}\frac{T}{T_{0}}\frac{\partial m^{2}}{\partial\epsilon} f(\bar{T}),
\end{align}

where 
\begin{equation}
     f(\bar{T},\bar{\Lambda})=\int d^{3}\overline{q}\int_{0}^{\infty}dx\frac{\left|\boldsymbol{\bar{q}}\right|x^{2}{\rm csch^{2}}\left(\frac{x}{2}\right)}{\left(x\bar{T}\right)^{2}+\boldsymbol{\bar{q}}^{2}\left(\bar{m}^{2}\left(\boldsymbol{q}\right)+\boldsymbol{\bar{q}}^{2}\right)^{2}}.
     \label{eq:f_efe_full}
\end{equation}
\end{comment}
\begin{align}
    &-\frac{\partial S}{\partial \epsilon}=-\frac{\partial F^2}{\partial \epsilon\partial T}\equiv -\gamma_0 \Gamma_{*,0}T f(\bar{T}),
\end{align}
where 
\begin{equation}
    \Gamma_{*,0}=-\frac{\Lambda^3}{32\pi^4m_*^2}\frac{\partial m^2}{\partial \epsilon}\frac{1}{\gamma_0 T_0},\label{eq:Gammastar0}
\end{equation}
as well as 
\begin{equation}
     f(\bar{T},\bar{\Lambda})=\int_0^{\bar{\Lambda}} d^{3}\overline{q}\int_{0}^{\infty}dx\frac{\left|\boldsymbol{\bar{q}}\right|x^{2}{\rm csch^{2}}\left(\frac{x}{2}\right)}{\left(x\bar{T}\right)^{2}+\boldsymbol{\bar{q}}^{2}\left(\bar{m}^{2}\left(\boldsymbol{q}\right)+\boldsymbol{\bar{q}}^{2}\right)^{2}}.
     \label{eq:f_efe_full}
\end{equation}
First, let us consider the limit of high temperatures $T \gg T_*$. At the quantum critical point, we can then neglect the $\bar{m}$ dependence of the denominator. Performing the momentum integral and subsequently the integral over $x$, the auxiliary function simplifies to
\begin{align}
     f(\bar{T},\bar{\Lambda})
     % =\int d^{3}\overline{q}\int_{0}^{\infty}dx\frac{\left|\boldsymbol{\bar{q}}\right|x^{2}{\rm csch^{2}}\left(\frac{x}{2}\right)}{\left(x\bar{T}\right)^{2}+\boldsymbol{\bar{q}}^{6}}
    %   &= \frac{4 \pi^2}{3\sqrt{3} \bar T^{2/3}} \int_{0}^{\infty}dx x^{4/3}{\rm csch^{2}}\left(\frac{x}{2}\right) \nonumber\\
     &= 32 \pi^4 b \frac{1}{\bar T^{2/3}} & {\rm for}\quad \bar{T}\gg 1.
\end{align}
where the numerical coefficient $b$ has been introduced below Eq.~\eqref{eq:dSde_bare}. This high-$T$ behaviour is consistent with the results obtained for the bare QCP in Ref.~\citep{Zhu2003}, see Eq.~\eqref{eq:dSde_bare}.

If the behaviour of the elastocaloric effect at low temperature $T \ll T_*$, were Fermi liquid-like, we could neglect the $\bar T$ dependence of the numerator in Eq.~\eqref{eq:f_efe_full}. Integrating over the amplitude of momentum would then result in an angular integral over the inverse squared mass, $\int d \hat q \frac{1}{m^2(\hat q)}$. In contrast to the case of the specific heat, this angular integral over the direction-dependent mass is logarithmically divergent due to the contributions of the soft directions. As a consequence, the behaviour is not consistent with Fermi liquid expectations, and we expect a residual logarithmic $T$-dependence that we extract in the following.

% We consider the same three regimes  of Eq.~\eqref{eq:rapprox} in momentum space as for the heat capacity.  We proceed first with the case ${\it (iii)}$ with $q_z\gg q_1,q_2$ such that $\bar{m}\left(\boldsymbol{q}\right)=1$. The integral can easily be analysed and yields
% \begin{align}
%  f^{(\it{iii})}(\bar{T},\bar{\Lambda})\approx \begin{cases}
%         \frac{8}{3}\pi^3, &\bar{T}\ll 1\\
%         32\pi^4 b \bar{T}^{-2/3}, & \bar{T}\gg 1
%     \end{cases}
%     ,
% \end{align}
% where the numerical coefficient $b$ was introduced below Eq.~\eqref{eq:dSde_bare}. We also verified this behaviour numerically. The low-temperature behaviour of a Fermi liquid obeying Gr\"uneisen scaling. Since this is the contribution from the gapped part of momentum space, the Fermi liquid behaviour is what one  expects. The high-$T$ power-law is consistent with the results obtained for the bare QCP in Ref.~\citep{Zhu2003}, see Eq.~\eqref{eq:dSde_bare}.  

This logarithmic dependence is attributed to the soft lines, i.e.,  
% Next we focus on the contributions from the soft lines, i.e, 
the two regimes {\it (i)} and {\it (ii)}. Both are again equivalent by symmetry and it suffices to analyse only one. Using the same spherical coordinates as before, we obtain
\begin{align}
    &f^{(\it{i})}(\bar{T},\bar{\Lambda})=f^{(\it{ii})}(\bar{T},\bar{\Lambda}) \nonumber \\
    &\approx 4\pi\int_{0}^{\overline{\Lambda}}dq\int_{0}^{\theta_{0}}d\theta\int_{0}^{\infty}dx\frac{\theta q^{3}x^{2}{\rm csch^{2}}\left(\frac{x}{2}\right)}{\left(x\bar{T}\right)^{2}+q^{2}\left(\theta^{2}+q^{2}\right)^{2}}.
\end{align}
In the low temperature limit this integral yields 
\begin{align}
f^{(\it{i})}(\bar{T},\bar{\Lambda})&\approx  \frac{8\pi^3}{9}\log\frac{\theta_{0}^3}{\bar{T}}, \quad \bar{T}\ll 1 
 \label{eq:fi_T}
\end{align}
% The analysis of  the integral in the low and high-$T$ limit is straightforward and yields
% \begin{align}
% f^{(\it{i})}(\bar{T},\bar{\Lambda})&\approx  \begin{cases}\frac{8\pi^3}{3}\log\frac{\theta_{0}}{\bar{T}}, & \bar{T}\ll 1 \\
%  16\pi^4\theta_{0}^{2}b\left(\frac{T}{T_{\ast}}\right)^{-2/3}, & \bar{T}\gg 1\end{cases},
%  \label{eq:fi_T}
% \end{align}
% To obtain this result we divide the integral into regions $0\leq\bar{q}\leq 1$, and $1\leq\bar{q}\leq \infty$. One finds  that the first region diverges as $T\rightarrow 0$ and gives the dominant contribution, while the other yields a constant $\frac{4\pi^3}{3}\log(1+\theta_0^2)$. Hence, the origin of the low-$T$ logarithmic  divergence are the gapless low energy excitations along the remaining soft lines in momentum space. At high-$T$ the second integral dominates and gives a behaviour in agreement with the bare critical point, while the first integral over the interval  $0\leq\bar{q}\leq 1$ decays faster as $T_\ast/T$.  Combining all the findings yields the result of Eq.~\eqref{eq:fi_T}. Notice, the two expressions don't match at $T\approx T_\ast$, a consequence of the fact that in the crossover regime, terms that are subleading at low and high-$T$ must be included.

Adding the contributions from two contributions $(i)$ and $(ii)$ in the low-$T$ limit, we obtain the result,
\begin{align}
   \frac{\partial S}{\partial\epsilon_{0}} &\approx -16\pi^3 \gamma_0\Gamma_{*,0}T_*
    \begin{cases}
        \frac{1}{9}\frac{T}{T_*}\log\frac{T_{\ast}}{T}, & T\ll T_{\ast}, \\[7pt]
        2\pi b\left(\frac{T}{T_{*}}\right)^{1/3}, & T\gg T_{\ast},
    \end{cases}
\end{align}
where, as before, the temperature scale $T_0$ for the bare and $T_\ast$ for the elastic QCP are given in Eqs. \eqref{eq:T_0} and  \eqref{eq:Tstar}, respectively, while the constant $\Gamma_{*,0}$ and numerical coefficient $b$ were introduced below Eq.~\eqref{eq:Gammastar0} and Eq.~\eqref{eq:dSde_bare}, respectively.  At high-$T$ we of course recover the bare QCP behaviour of Eq.~\eqref{eq:dSde_bare}, while the new result is the logarithmic dependence at low $T$ that is due to soft lines in momentum space. Notice, the two expressions don't match at $T\approx T_\ast$ because in the crossover regime subleading terms in low and high-$T$ must be included.

Using the  results of Eq.~\eqref{eq:Sommerfeld_el} for the Sommerfeld coefficient and the above results, we can finally obtain the elastocaloric effect as:
\begin{align}
    \eta &\approx -72 \Gamma_{*,0}T_* 
    \begin{cases}
        \frac{1}{9} \frac{\left(T/T_{\ast}\right)\log\frac{T_{\ast}}{T}}{\log\frac{T_{0}}{T_{\ast}}}, & T\ll T_{\ast}, \\[7pt]
        2\pi b \frac{\left(T/T_{*}\right)^{1/3}}{\log\frac{T_{0}}{T}}, & T\gg T_{\ast}.
    \end{cases}
\end{align}

\begin{figure}
    \centering
    \includegraphics[width=0.9\linewidth]{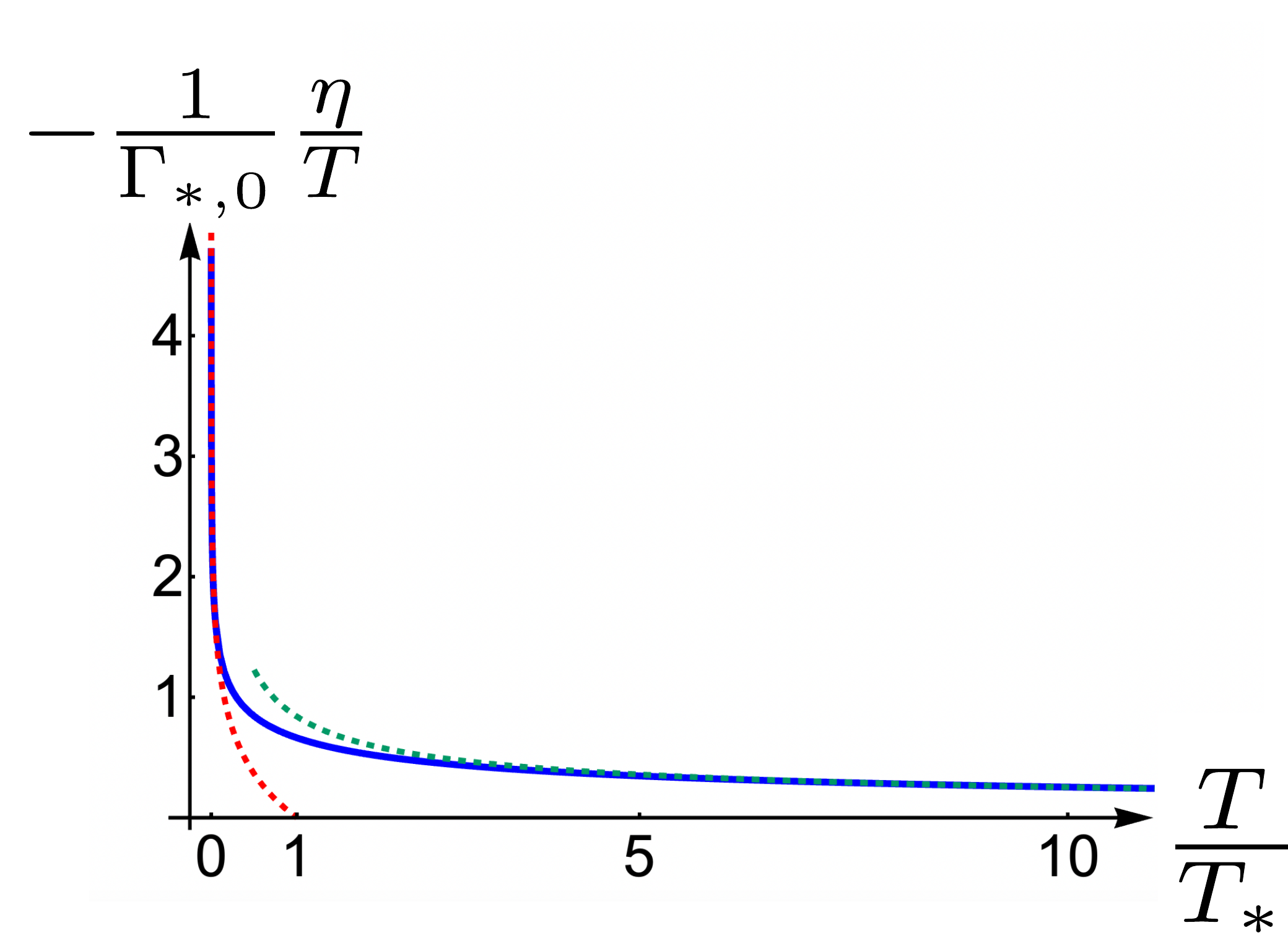}
    \caption{Elastocaloric coefficient $\eta$ divided by $T$, i.e.\ the Gr\"uneisen parameter,  as a function $T/T_{\ast}$ at the elastic quantum critical point,  from the numerical analysis of Eq.~\eqref{eq:f_efe_full}. At low temperature $(T\ll T_\ast)$ it shows the $\log T$ behaviour as indicated by the red dashed lines, while in the high temperature regime ($T\gg T_\ast$) it shows the $T^{-2/3}/\log(T_0/T)$ behaviour as indicated by the green dashed line. The rather broad crossover regime, mentioned in the text is clearly visible.}
    \label{fig:ElastoCaloricChageAtElasticQC}
\end{figure}

This is the behaviour listed in Tab.~\ref{tab:tab01}. As mentioned earlier,   the two expressions do not perfectly match at $T=T_{\ast}$; subleading  terms would need to be considered to model the $T\sim T_{\ast}$ crossover behaviour. This can be seen clearly from the numerical simulation shown in  Fig.~\ref{fig:ElastoCaloricChageAtElasticQC}, where we observe a broad crossover regime. Most importantly, in agreement with our analytic analysis, the low temperature behaviour of $\eta$ does not follow the Fermi liquid theory, even if the coupling to strain is taken into consideration.

In section~\ref{sec:scaling} we discuss that this distinct behaviour is due to the fact that critical contributions to the heat capacity are subleading, while they are the dominant one for the elastocaloric effect.

As a result, the elastocaloric change in temperature is a good measure for the direction-selective gapless mode of the boson due to nematoelastic coupling. The difference in the elastocaloric effect when above or below $T_{\ast}$ suggests that such an experiment would allow one to deduce whether the system is in the elastic QC or bare QC regime. In iron-based superconductors, the crossover temperature has been estimated to be 
%approximated at 
$10\si{K}$~\cite{Paul2017}, which suggests that both regimes are accessible in lab conditions. The full phase diagram showing the bare quantum critical and elastic quantum critical regimes is shown in Fig.~\ref{fig::phase_01}.

The free energy expression of Eq.~\eqref{eq:free_eng} is the contribution due to the
bosonic nematic excitations whose propagator $\chi\left(\omega,\boldsymbol{q}\right)$
is modified by the coupling to fermions and strain. We could alternatively
write this in terms of the coupling constant integration~\cite{abrikosov2012methods}
\begin{equation}
F=F_{0}-\int_{0}^{1}\frac{d\lambda}{\lambda}T\sum_{\omega}\int\frac{d^{3}q}{\left(2\pi\right)^{3}}\Pi\left(\omega,\boldsymbol{q}\right)\chi\left(\omega,\boldsymbol{q}\right),
\end{equation}
where $F_{0}$ is the free energy at $g=0$ and $\Pi\left(\omega,\boldsymbol{q}\right)$
the bosonic self energy that is responsible for the Landau damping
in $\chi\left(\omega,\boldsymbol{q}\right)$. As usual we replaced
the coupling constant $g$ by $\lambda g$ and consider the solutions
for different $\lambda$. The resulting heat capacity expressions
of Eqs.~\eqref{eq:gamma_0} and \eqref{eq:gamma0}, with and without the coupling to strain, respectively
are identical to the ones of a purely fermionic system with self energy
$\Sigma\left(\omega,\boldsymbol{k}\right)$ and Green's function $G\left(\omega,\boldsymbol{k}\right)$.
This is to be expected as the same free energy can alternatively be
written as~\cite{abrikosov2012methods}
\begin{equation}
F=F_{0}+2\int_{0}^{1}\frac{d\lambda}{\lambda}T\sum_{\omega}\int\frac{d^{3}k}{\left(2\pi\right)^{3}}\Sigma\left(\omega,\boldsymbol{k}\right)G\left(\omega,\boldsymbol{k}\right),
\end{equation}
reflecting the fact that the corrections to the free energy due to
electron-boson interaction cannot be allocated to only one of the
two coupled degrees of freedom. Hence, the same anomaly that is responsible
for the deviation from Fermi liquid behavior of the elastocaloric effect is equally
present in the electronic self energy. While the underlying non-analytic
corrections are sub-leading for the self energy itself, they become
dominant if one considers strain derivatives that are important for
the elastocaloric effect. Analyzing these effects from the bosonic perspective, as done in our theory, is merely a choice of convenience.

\section{Piezomagnetic coupling in altermagnets}
\label{sec:StrainCouplingAlt}
There exists an interesting resemblance between the nematoelastic coupling Eq.~\eqref{eq:nemel} and the piezomagnetic coupling of Eq.~\eqref{eq::Field_coupling} in altermagnets. The coupling of elastic degrees of freedom to a nematic order parameter has been discussed in some detail in the literature~\citep{Cowley1976,Karahasanovic2016,Paul2017} and is primarily based on the interaction term given in Eq.~\eqref{eq:nemel}. Hence, in this section we discuss the strain coupling of altermagnets, focusing on the group theory principles which underpin Eq.~\eqref{eq::Field_coupling}.

Consider an altermagnet, with order parameter $\phi^a$
that corresponds to multipolar magnetism, i.e., it breaks time-reversal
symmetry, keeps translation invariance, and does not transform like
a magnetic dipole. The latter ensures that the net moment in the unit cell is zero by symmetry. The index $a$ labels the components of the order parameter; this index we drop when we consider single-component altermagnets.
We denote the irreducible representation of the
order parameter as $\Gamma_\phi^{-}$, where the minus sign indicates that
the representation is odd under time-reversal. For a $g$-dimensional representation, $a=1,\ldots,g$ labels the components.
The symmetry-allowed coupling of $\phi^a$ to lattice fluctuations will be crucial for our subsequent analysis.

As discussed in Ref.~\citep{Steward2023},
there are multiple options for the order parameter to couple to the
lattice. On the one hand, a dynamic strain field that transforms
under $\Gamma_\phi^{+}$ can couple to the conjugate momentum of $\phi^a$.
This  coupling gives rise to important modifications of the
phonon and altermagnon spectrum at finite frequencies and modifies
the region in the phase diagram where altermagnetism can be stabilized~\citep{Steward2023}. In particular, this coupling  does not change the universality class of critical
fluctuations of the order parameter. On the other hand, there is the piezomagnetic coupling of 
Eq.~\eqref{eq::Field_coupling}. 
%This is true for  both,  classical fluctuations at finite temperatures and quantum fluctuations near a putative quantum critical point.
To make this interaction term more tangible, we determined for several important crystalline point groups the allowed magnetic field-induced coupling to strain, along with
the required direction of the magnetic field and the symmetry of the strain fields. The piezomagnetic coupling of Eq.~\eqref{eq::Field_coupling}
allows probing the altermagnetic
order parameter by measuring the strain dependence of
the magnetization 
\begin{equation}
   \sum_{a} \lambda^a_{\alpha,i} \phi^a=\partial M_{\alpha}/\partial\epsilon_{\Gamma_{i}^{+}}.
\end{equation}
Choosing the appropriate field direction and strain symmetry enables us to probe and determine  the otherwise rather hidden altermagnetic order parameter. 
The symmetry allowed  piezomagnetic couplings for a tetragonal system is listed in Table~\ref{Tab1}. In addition we summarize the couplings for systems with orthorhombic (Table~\ref{Tab4}), hexagonal (Table~\ref{Tab2}) and cubic (Table~\ref{Tab3}) symmetries in the Appendix~\ref{sec::softening}. 
In these tables  we also list the symmetry-allowed coupling of $\phi^a$  to fermionic modes of a single band system. The latter was discussed in the context of protected nodal lines of the Zeeman splitting~\citep{fernandes2024topological}, and will be crucial when we analyse the dynamics near altermagnetic quantum critical points.
%We will demonstrate that the coupling of Eq.~\eqref{eq::Field_coupling} allows for a field-tunable change of the universal critical
%properties of altermagnetic fluctuations. 

%is also the origin of  response of altermagnets.
%The uniform magnetization density
%\begin{equation}
%M_{\alpha}=-\frac{1}{V}\left.\frac{\partial F}{\partial H_{\alpha}}\right|_{H\rightarrow0}=\sum_{i,a}\lambda^a_{\alpha,i}\epsilon_{\Gamma_{i}^{+}}\phi^a,
%\label{eq::piezomag}
%\end{equation}
%can be induced via symmetry-breaking strain. 

The classification of Tab.~\ref{Tab1}-\ref{Tab3} reveals several aspects that are worth summarizing. First, the piezomagnetic coupling of Eq.~\eqref{eq::Field_coupling} is not limited to altermagnetic states but equally occurs in some systems with ferromagnetic order. Those are indicated in gray in the tables. Applying the field along a hard direction of a ferromagnet allows for  behaviour similar to the one in altermagnets.  Second, in some cases, such as tetragonal crystals, piezomagnetic couplings can be used to uniquely identify the symmetry of an altermagnetic state, while for other symmetries, such as cubic and hexagonal systems, this is not the case. Finally, there also exist altermagnetic states that cannot be detected through piezomagnetic couplings. An example is an order parameter that transforms under $A_{1g}^-$ 
in cubic crystals; see Table~\ref{Tab3}. This state is a combination of a spin dipole and a charge hexadecapole that forms a dotriacontapolar state. It can, however, be identified if one simultaneously applies strain of $T_{2g}^+$ and $E_g^+$ symmetries and analyses piezomagnetic effects that are nonlinear in the strain fields.

The piezomagnetic coupling, Eq.~\eqref{eq::Field_coupling} allows one to make some analogies to the behaviour in nematic systems, where the order parameter also breaks a point group symmetry, but is even under time reversal. Hence if we absorb the magnetic field in the coupling constant 
\begin{equation}
\lambda_{H}=\sum_{\alpha}\lambda_{\alpha,i}H_{\alpha},
\end{equation}
the system essentially behaves as a nematic system, yet with different symmetries of the strain field, as dictated by Tab.~\ref{Tab1}-\ref{Tab3}. 
If we then follow the analysis of Ref.~\citep{fernandes2010effects} for nematic systems, one finds a relation identical to Eq.~\eqref{eq::elasticrenorm_nem} but in which $\lambda$ is replaced by $\lambda_H$ and the nematic susceptibility $\chi_{{\rm nem}}$ is replaced by the altermagnetic  susceptibility $\chi_{{\rm am}}$.
%\begin{equation}
%C^{-1}=C_{0}^{-1}+\frac{\lambda_{H}^{2}}{C_{0}^{2}}\chi_{{\rm am}}.\label{eq::elasticrenorm}
%\end{equation}
 The divergence of the altermagnetic susceptibility at a second order phase transition gives rise to a field-induced lattice softening. The latter can then be used to measure, at least in principle, the 
order parameter susceptibility $\chi_{\rm am}$. In practice, the small energy scale of the field, compared to the typical energy scales of the ordering temperature, makes these effects hard to observe in currently known altermagnets. %We provide details on the derivation of Eq.~\eqref{eq::elasticrenorm} in Appendix~\ref{sec::softening}. 

\begin{table}
{\renewcommand{\arraystretch}{1.3}
\renewcommand{\tabcolsep}{3.3pt}
\begin{tabular}{c|c|c}
\hline \hline
\multicolumn{3}{c}{\normalsize $D_{4h}$ ($4/mmm$) point group} \tabularnewline
AM irrep & coupling to fermions & piezomagnetic coupling \tabularnewline
\hline
$A_{1g}^{-}$ & $g \phi k_z \! \left(k_y\sigma_x - k_x \sigma_y\right)$ & $\lambda\phi\left(\epsilon_{yz}H_x - \epsilon_{xz} H_y\right)$ \tabularnewline
\hline 
$B_{1g}^{-}$ & \begin{tabular}{c}
$g \phi k_z \! \left(k_y \sigma_x + k_x \sigma_y\right)$\tabularnewline
$g' \phi k_x k_y \sigma_z$\tabularnewline
\end{tabular} & \begin{tabular}{c}
$\lambda\phi\left(\epsilon_{yz} H_x + \epsilon_{xz} H_y\right)$\tabularnewline
$\lambda'\phi\epsilon_{xy}H_z$\tabularnewline
\end{tabular} \tabularnewline
\hline 
$B_{2g}^{-}$  & \begin{tabular}{c}
$g \phi k_z \! \left(k_x \sigma_x - k_y \sigma_y\right)$\tabularnewline
$g' \phi\left(k_x^2-k_y^2\right)\sigma_z$\tabularnewline
\end{tabular} & \begin{tabular}{c}
$\lambda\phi\left(\epsilon_{xz} H_x - \epsilon_{yz} H_y\right)$\tabularnewline
$\lambda'\phi\epsilon_{x^2-y^2}H_z$\tabularnewline
\end{tabular} \tabularnewline

\hline
FM irrep & & \tabularnewline
\hline
\textcolor{FMirrepColour}{$A_{2g}^{-}$} & \textcolor{FMirrepColour}{$g \phi \sigma_z$} & \begin{tabular}{c}
\textcolor{FMirrepColour}{$\lambda\phi\epsilon_{A_{1g}}H_z$}\tabularnewline
\textcolor{FMirrepColour}{$\lambda'\phi\left(\epsilon_{zx}H_x + \epsilon_{yz}H_y\right)$}
\end{tabular}\tabularnewline
\hline 
$\textcolor{FMirrepColour}{E_{g}^{-}}$ & \textcolor{FMirrepColour}{$g \left(\phi_1 \sigma_x + \phi_2 \sigma_y\right)$} & \begin{tabular}{c}
\textcolor{FMirrepColour}{$\lambda\epsilon_{A_{1g}}\left(\phi_1 H_x + \phi_2 H_y\right)$}\tabularnewline
\textcolor{FMirrepColour}{$\lambda'\epsilon_{x^2-y^2}\left(\phi_1 H_x - \phi_2 H_y\right)$}\tabularnewline
\textcolor{FMirrepColour}{$\lambda''\epsilon_{xy}\left(\phi_1 H_y + \phi_2 H_x\right)$}\tabularnewline
\textcolor{FMirrepColour}{$\lambda'''\left(\phi_1\epsilon_{xz}+\phi_2\epsilon_{yz}\right)H_z$}
\end{tabular}\tabularnewline
\hline \hline
\end{tabular}}
\caption{The coupling of altermagnetic (AM) and ferromagnetic (FM) order parameters $\phi$ to fermions and to simultaneous magnetic and strain fields (piezomagnetism) for the tetragonal point group $D_{4h}$ ($4/mmm$).
Ferromagnetic $\phi$ have been listed for the sake of completeness.
The first column indicates the irreducible representation (irrep) according to which $\phi$ transforms.
For $E_g$, $(\phi_1,\phi_2)$ transforms the same as $(\sigma_x,\sigma_y)$.
$g$ and $\lambda$ are coupling constants, $k_{\alpha}$ is the momentum, $\sigma_{\alpha}$ are Pauli matrices, $\epsilon_{\alpha\beta}$ is the strain tensor, and $H_{\alpha}$ is the magnetic field.
The possibility of having $g' \neq g$ and $\lambda' \neq \lambda$ reflects magnetic anisotropy.
$\epsilon_{A_{1g}}$ are the trivially transforming strain components.
The coupling to lattice fermions is deduced by replacing $k_{\alpha}$ with $\sin k_{\alpha}$.
The results of this and other tables agree with Ref.~\citep{fernandes2024topological} whenever they overlap.}
\label{Tab1}
\end{table}

\section{Elastic quantum criticality at an altermagnetic QCP}
\label{sec:altQCP}
Nematic and altermagnetic order parameters both transform according to a similar set of spatial transformations and mainly differ in their behaviour under  time-reversal. This suggest that a similar analysis to strain coupling can be carried out for elastic quantum criticality at an altermagnetic QCP in the presence of an external magnetic field. For a $B_{2g}$ altermagnet in an arbitrary magnetic field, the coupling term is according to Table~\ref{Tab1} given by
 \begin{equation}
     \mathcal{S}=\int_x\phi\left(x\right)\left(\lambda_z H_z\left(\epsilon_{xx}\left(x\right)-\epsilon_{yy}\left(x\right)\right)+\lambda_{\perp}\left(\epsilon_{xz} H_x - \epsilon_{yz} H_y\right)\right),
 \end{equation}
 where $\lambda_z$ is the coupling to $B_{1g}$ strain, mediated by magnetic field in the $z$ direction and $\lambda_{\perp}$, the coupling to $E_{1g}$ strain, mediated by the in-plane magnetic field, respectively. It is crucial that the field only modifies the properties of the QCP, but does not destroy it, as would be the case for a ferromagnetic transition with a field along the easy direction.
 
\begin{figure}
     \centering
     \includegraphics[width=0.48\textwidth]{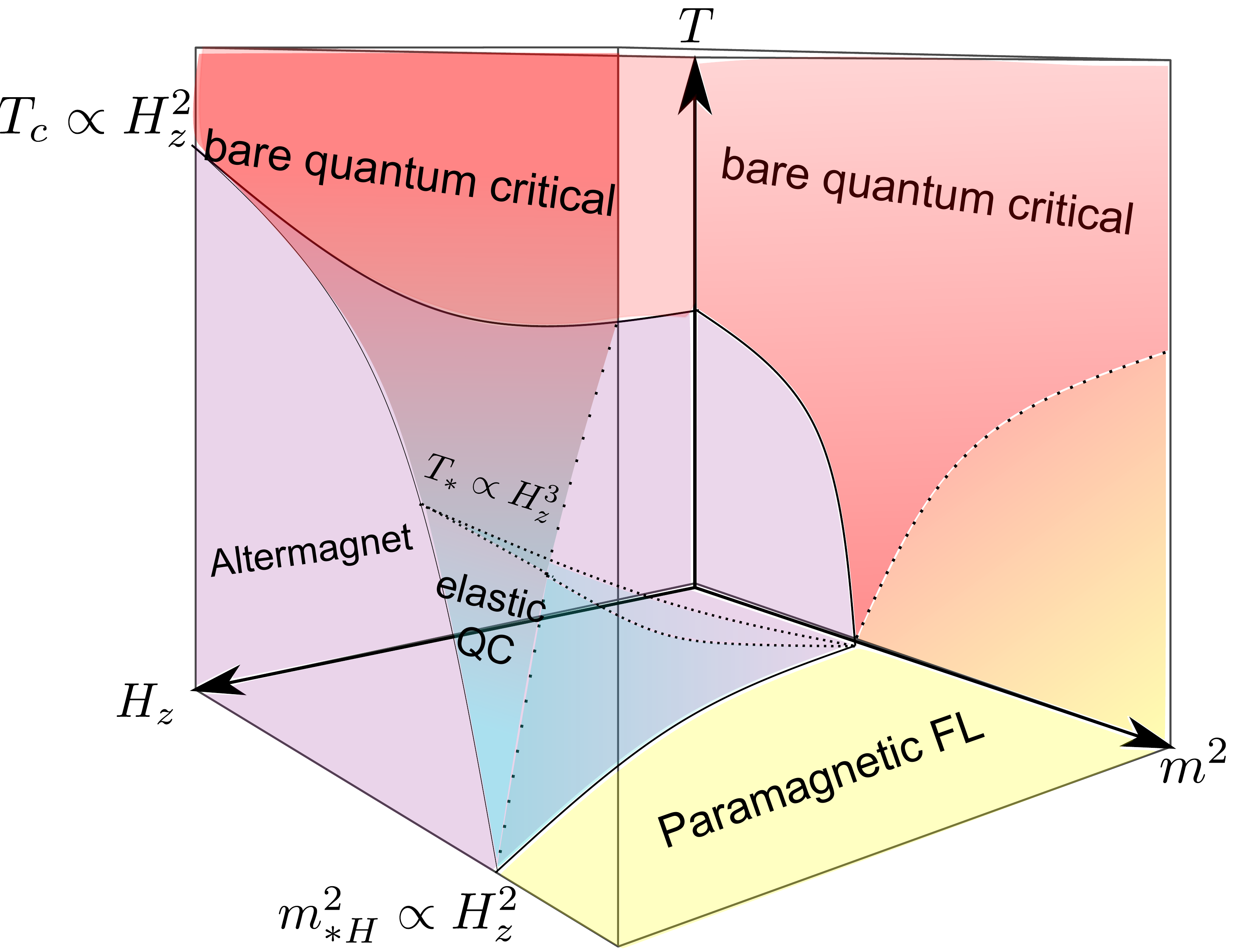}
     \caption{The full phase diagram for an altermagnet in a magnetic field shows initially at zero field an altermagnetically ordered regime and a disordered paramagnetic FL phase meeting at a QCP, around this point is a region of NFL behaviour. Turning on the magnetic field increases the area of the ordered regime and introduces a new region of FL behaviour around the QCP, distinct from the paramagnetic phase. Phase changes are marked with solid lines and crossover behaviour with dotted lines.}
     \label{fig::3Dphase}
 \end{figure}

The main difference, compared to a nematic system, is a field-tunable coupling strength $\lambda_H\sim \lambda_z H_z $. In addition,   we will  see that the time-reversal odd nature of the altermagnet plays an interesting role for the Landau damping. The condition on the  Fermi surface for Landau damping is automatically fulfilled.  The role of the field tuning for the quantum critical regime  is summarized in Fig.~\ref{fig::3Dphase}. In addition to the tuning by some non-thermal parameter, the quantum critical point varies with field allowing one, in principle, to reach a QCP by increasing an appropriate component of $\mathbf{H}$. Then, the QCP should, however, be governed by elastic quantum criticality.
 
\subsection{Direction selective criticality}
 We can, in analogy to the nematic system, integrate out the elastic modes and obtain for the $B_{2g}$ altermagnet in a field, the direction-dependent correction to the order parameter action:
\begin{align}
\Delta \mathcal{S}^{({\rm am})}_{\epsilon-\phi}&=-\frac{1}{8}\int_q\phi(q)\phi(-q)\Bigg[\frac{\lambda_\perp^2(H_x q_x-H_yq_y)^2}{\rho\Omega^2_1(\mathbf{q})}\nonumber\\
&+\frac{1}{|\mathbf{q}_{2d}|^2}\frac{\Big(\lambda_zH_z(q_x^2-q_y^2)-\lambda_\perp q_z(H_xq_x-H_yq_y)\Big)^2}{\rho\Omega^2_2(\mathbf{q})}\nonumber\\
&+\frac{1}{|\mathbf{q}_{2d}|^2}\frac{\Big(2\lambda_zH_zq_xq_y-\lambda_\perp(H_yq_x+H_xq_y)q_z\Big)^2}{\rho\Omega_3^2(\mathbf{q})}\Bigg],
\end{align}
with the same elastic theory as for the nematics. Choosing $\mathbf{H}=H\hat{z}$ gives a near-identical coupling term as for the $B_{1g}$ nematic but with a field-tuned coupling strength.
The mass is renormalized and becomes 
\begin{align}
m^2(\mathbf{q},H_z)&=m^2- \frac{(\lambda H_z)^2}{4\rho}\frac{1}{|\mathbf{q}_{2d}|^2}\Big(\frac{(q_x^2-q_y^2)^2}{\Omega^2_2(q)}+\frac{4q_x^2q_y^2}{\Omega^2_3(q)}\Big).\label{eq::RenormalizedMassHzNeqZeroFull}
\end{align}
In analogy with the nematic case, $m^2(\mathbf{q},H_z)$ has minimum value 
\begin{align}
    m^2_{\rm min}(H_z)=m^2-m^2_{\ast H},\label{eq:renormalizedMassAltermagnet}
\end{align}
where $m^2_{\ast H}=\frac{\lambda^2H_z^2}{2\left(C_{11}-C_{12}\right)}$, along the two lines $q_x=\pm q_y$. We see that by varying the magnetic field strength, the position of the critical point is also moved, i.e., a transition can be  field-induced; see Fig.~\ref{fig::zeroT}. The renormalized critical mass ($ m_{\rm min}=0$) approximates near the high symmetry directions to
\begin{align}
    m^2(\mathbf{q},\mathbf{H})\approx m^2_{\ast H}\left\{ \begin{array}{ll}\frac{ q_2^2+ q_z^2}{q_1^2}, & |q_1|\gg |q_2|,|q_z|\\
    \frac{ q_1^2+ q_z^2}{q_2^2}, & |q_2|\gg |q_1|,|q_z|\\
    1, & |q_z|\gg |q_1|,|q_2|,
    \end{array}\right.\label{eq::nonZeroHzRenormalizedMass}
\end{align}
where $q_1=\frac{q_x+q_y}{\sqrt{2}}$ and $q_2=\frac{q_x-q_y}{\sqrt{2}}$. Here we set the coefficients to be 1 for simplicity. Note that we use the notation $m^2_{\ast H}$ instead of $m^2_{\ast}$, which is used in nematic case, which shows that the $m^2_{\ast H}$ is dependent on the magnetic field at criticality for the altermagnetic case.  We will see that this relation implies a magnetic field dependent crossover temperature $T_{\ast}$ proportional to $H_z^3$. In our subsequent analysis we discuss  the behaviour for a spherical Fermi surface. Results for a cylindrical Fermi surface are given in Appendix~\ref{App:CylindricalFS}.

\subsection{Landau damping}\label{sec::alter_Landau}
The coupling of electrons to the order parameter is given by
\begin{align}
\mathcal{S}^{({\rm am})}_{c-\phi}&=\int_{k,q}\phi(q) c_\alpha^\dagger(k+q/2) \Big[g_1h_z(k)\sigma^z_{\alpha\beta}\nonumber\\
&+g_2\Big(h_x(k)\sigma^x_{\alpha\beta} +h_y(k) \sigma^y_{\alpha\beta}\Big)\Big]c_\beta (k-q/2),
\end{align}
 where it follows with the help of Table~\ref{Tab1} that
 \begin{eqnarray}
   \mathbf{h}(k)&=&(h_x(k),h_y(k),h_z(k)) \nonumber \\
   &=&(-\sin k_x\sin k_z,\sin k_y\sin k_z,\cos k_x -\cos k_y).
 \end{eqnarray}
The spin  Pauli matrices are a consequence of the broken time-reversal symmetry.
The one-loop correction to the order-parameter propagator due to the fermion-boson coupling is given by  
\begin{eqnarray}
\Delta \mathcal{S}_\phi^{c}
&=&\frac{1}{2}\int_q \phi(q)\phi(-q)\nonumber \\
&\times &\Big[g_1^2D_z(q)+g_2^2\Big(D_x(q)+D_y(q)\Big)\Big],
\end{eqnarray}
where $tr[\sigma^i\sigma^j]=2\delta^{ij}$ was used, while  
\begin{align}
D_i(q)&=2\int_k h_i^2(\mathbf{k})G(k-q/2)G(k+q/2)\nonumber \\
&=-2i\int\frac{d^3 k}{(2\pi)^3}h_i^2(\mathbf{k})\frac{\theta(\xi_{\mathbf{k}-\mathbf{q}/2})-\theta(\xi_{\mathbf{k}+\mathbf{q}/2})}{\Omega+i(\xi_{\mathbf{k}-\mathbf{q}/2}-\xi_{\mathbf{k}+\mathbf{q}/2})}. \label{eq::LandauDampingFormula}
\end{align}
Here, $i=x,y,z$ and $q=(\Omega,\mathbf{q})$. 

\begin{figure}[t]
    \centering
    \includegraphics[width=0.95\linewidth]{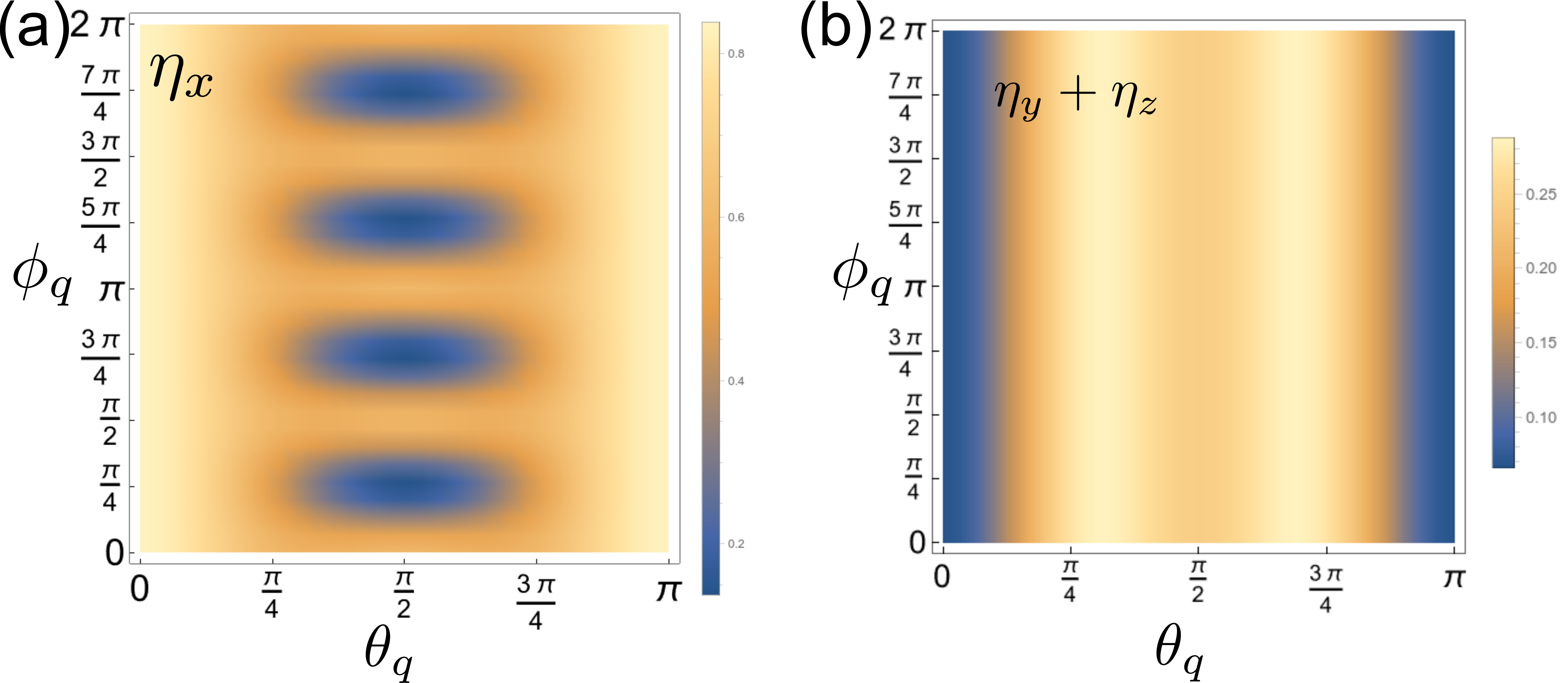}
    \caption{Panel (a) shows that $\eta_x$ has a term linear in $\Omega$ everywhere, except when $\theta_q=0$ and $\phi_q=\left(2n+1\right)/4\pi$, with $n=0,1,2,3$, where the damping is approximately quadratic in $\Omega$. A density plot of $\eta_y+\eta_z$ shows a term linear in $\Omega$ when $\theta_q= 0,\pi$. This suggests that when the two are summed, the main contribution comes from the linear $\Omega$ term at low energies everywhere on the Fermi surface. This is plotted at $\varphi=0.1$}
    \label{fig::landau_damping}
\end{figure}

With the assumption of a spherical Fermi surface ($\xi_{\mathbf{k}}=\frac{|\mathbf{k}|^2}{2m}-\mu$) and a low temperature approximation, we can obtain following expression for $D_i(q)$
\begin{align}
    D_i(q)\approx \frac{2i}{v_{\rm{F}}}\int\frac{d^3k}{(2\pi)^3}h_i^2(k)\delta(k-k_{\rm{F}})\frac{v_{\rm{F}}\mathbf{q}\cdot\hat{k}}{\Omega-iv_{\rm{F}}\mathbf{q}\cdot\hat{k}},
\end{align}
where $v_{\rm{F}}$, $k_{\rm{F}}$ and $\hat{k}$ are the Fermi velocity, wave vector and the normalized radial vector respectively. 
To determine the fermion-induced dynamics, we introduce  $\Delta D_i(q)$ given by 
\begin{align}
    \Delta D_i(q)&\equiv D_i(\Omega,\mathbf{q})-D_i(0,\mathbf{q})\nonumber\\
    &=\frac{2}{v_{\rm{F}}}\int \frac{d^3k}{(2\pi)^3}h_i(\mathbf{k})\delta(k-k_{\rm{F}})\frac{1}{1-iv_{\rm{F}}\mathbf{q}\cdot\hat{k}/\Omega}.
\end{align}
The term $D_i(0,\mathbf{q})$ modifies $m$ and gives a $|q|^2$ term which is related to the bare kinetic term of the boson.
Using spherical coordinates we can obtain the following results:
\begin{align}
    \Delta \mathcal{S}_\phi^c & \approx \int_q \frac{1}{2}\phi(q)\frac{k_{\rm{F}}^2(k_{\rm{F}}a)^4}{4\pi^3v_{\rm{F}}}\Bigg[g_1^2\eta_x(\theta_q,\phi_q,\varphi)\nonumber \\
    &+g_2^2\Big(\eta_y(\theta_q,\phi_q,\varphi)+\eta_z(\theta_q,\phi_q,\varphi)\Big)\Bigg]\phi(-q),
\end{align}
where $\theta_q$, $\phi_q$ are the polar  and the azimuthal angles of the momentum $\mathbf{q}$, respectively, while  $\varphi=\frac{|\Omega|}{v_{\rm{F}}|\mathbf{q}|}$. The functions $\eta_x$, $\eta_y$ and $\eta_z$ are given by 
\begin{align}
\eta_x&=\int\frac{ d\theta d\phi\sin^5\theta\cos^2(2\phi)}{1+\varphi^{-2}[\cos\theta\cos\theta_q+\cos(\phi-\phi_q)\sin\theta\sin\theta_q]^2},\\
\eta_y&=\int  \frac{ d\theta d\phi\sin^3\theta \cos^2\theta \cos^2\phi}{1+\varphi^{-2}[\cos\theta\cos\theta_q+\cos(\phi-\phi_q)\sin\theta\sin\theta_q]^2},\\
\eta_z&=\int \frac{ d\theta d\phi\sin^3\theta \cos^2\theta \sin^2\phi}{1+\varphi^{-2}[\cos\theta\cos\theta_q+\cos(\phi-\phi_q)\sin\theta\sin\theta_q]},
\end{align}
where we have suppressed the $\theta_q$, $\phi_q$ and $\varphi$ dependence of the $\eta$'s and the integrals are taken over the unit sphere.
Performing a numerical analysis, we find that there always exists a damping term regardless of the Fermi surface geometry, which we show in Fig.~\ref{fig::landau_damping}; this is in direct contrast with the nematic case. We approximate this damping term in the low energy limit as
    \begin{align}
    \Delta \mathcal{S}_\phi^c\approx \frac{6\pi\alpha v_{\phi}^{2}k_{F}^{2}}{2} \int\frac{d^4q}{(2\pi)^4}\phi(q)\frac{|\Omega|}{v_{\rm{F}}|\mathbf{q}|}\phi(-q)\label{eq::DeltaSphiFromPhi}.
\end{align}
We note that this result is dependent on the shape of the Fermi surface, as Landau damping is possible due to the creation of particle-hole pairs where the soft directions characterising the critical regions are parallel to the Fermi surface~\cite{Paul2017}, while these results naturally apply to other Fermi surface geometries (Appendix~\ref{App:CylindricalFS}), an exception occurs for nested Fermi surfaces or the behavior in the
vicinity to Van Hove points, where our theory does not apply.

 \subsection{Heat capacity}\label{sec::alter_heat}
Following the same procedure as in Sec.~\ref{sec::nem_heat},  we obtain for the Sommerfeld coefficient
%To calculate the contribution to the specific heat from the boson particle, we use the following one-loop dressed boson dispersion.
%\begin{align}
%\chi^{-1}(\Omega,\mathbf{q})&\approx c^2|\mathbf{q}|^2+r(\mathbf{q},\mathbf{H})+\nu_0\textcolor{red}{\zeta}(\theta_q,\phi_q) \frac{|\Omega|}{v_{\rm{F}}|\mathbf{q}|},
%\end{align}
% and the term $\Omega^2$ in the original boson action is ignored since the Landau damping term becomes dominant in the low energy limit. $r(\mathbf{q},\mathbf{H})$ is the mass which has been renormalized by the elastic coupling (Eq.~\eqref{eq::RenormalizedMassHzNeqZeroFull}). For simplicity, we ignored the angle-dependent Landau damping term (i.e.\ $\zeta(\phi_q,\phi_q)\approx \zeta$) since there is always an angle-independent damping term such that there is damping present everywhere on the Fermi surface. 

%Additionally, assuming that the bare dispersion of the boson comes from the fermion-boson interaction, we can show that $\frac{c^2}{\nu_0\zeta}\sim \frac{1}{k_{\rm{F}}^2}$. As a result,
%\begin{align}
 %   \chi^{-1}(\Omega,\mathbf{q},\mathbf{H})\propto \frac{|\mathbf{q}|^2}{k_{\rm{F}}^2}+\tilde{r}(\mathbf{q},\mathbf{H})+\frac{|\Omega|}{v_{\rm{F}}|\mathbf{q}|},\label{eq::OneLoopBosonDispersion}
%\end{align}
%where $\tilde{r}=\frac{r}{\nu_0\zeta}$. In the remaining text, we will use $r(\mathbf{q},\mathbf{H})$ to denote $\tilde{r}(\mathbf{q},\mathbf{H})$ for simplicity. 

%Using the above $\chi^{-1}(\Omega,\mathbf{q})$, we obtain the following general expression of the specific heat

\begin{align}
    \gamma(T,\mathbf{H})&=\gamma_0\int d^3\bar{q}\int_0^\infty dx \frac{x}{e^x-1}\nonumber\\
    &\times \frac{|\bar{\mathbf{q}}|^3\Big(|\bar{\mathbf{q}}|^2+\bar{m}^2(\bar{\mathbf{q}},\mathbf{H})\Big)^3}{\Big[|\bar{\mathbf{q}}|^2\Big(|\bar{\mathbf{q}}|^2+\bar{m}^2(\bar{\mathbf{q}},\mathbf{H})\Big)^2+x^2\bar{T}^2\Big]^2},\label{eq::GeneralExpressionForSpecificHeat}
\end{align}
where $\gamma_0$ is introduced in Eq.~\eqref{eq:gamma_0} and the dimensionless parameters $\bar{\mathbf{q}}$, $\bar{T}$ and $\bar{m}$ are given by the same expressions as in Eq.~\eqref{eq:dimless_para}, yet with a field-dependent $m^2_{\ast H}$.
For the spherical Fermi surface considered here, there are no qualitative dependencies of the specific heat  on the direction of the magnetic field. Therefore we consider the $\mathbf{H}=H\hat{z}$ case only. We also confirmed that, for a  cylindrical Fermi surface the specific heat shows qualitatively the same behaviour as for the spherical Fermi surface case regardless of the direction of magnetic field. 

Elastic quantum criticality corresponds  $m_{\rm min}=0$, see Eq.~\eqref{eq:renormalizedMassAltermagnet}. In what follows we  use the Eq.~\eqref{eq::nonZeroHzRenormalizedMass} for $m(\mathbf{q},\mathbf{H})$.   From an analysis that proceeds in full analogy to the case of a nematic QCP, we obtain
\begin{align}
    \gamma(T)\approx\frac{2\pi^3}{9}\gamma_0\begin{cases}
    \log \frac{T_0}{T_\ast}\propto -\log |H_z|, & T\ll T_{\ast}, \\[7pt]
    \log \frac{T_0}{T}, & T\gg T_{\ast}.
    \end{cases}
    \label{eq::nonZeroHSpecificHeat}
\end{align}
where $T_{\ast}\sim \frac{\lambda^{3}H_z^3}{C^{3/2}J^{3}}T_{0}$. Note that $T_\ast\propto H_z^3$ since $m^2_{\ast H}\propto H_z^2$ at the critical point. Hence, the crossover scale can be tuned by changing the magnetic field; see Fig.~\ref{fig::3Dphase}. Overall, the Sommerfeld coefficient of an altermagnet in the field behaves similar to a nematic system.

\subsection{Electron self-energy}
These results are again supported by a calculation of the electron self-energy, which is also changed slightly as a result of the expanded form factor. We carry out this calculation to one-loop order. The self-energy of an electron exchanging bosons is given by
\begin{equation}
\Sigma\left(\mathbf{k},\omega_n\right)\propto h_{k}^2\int_{q,\Omega_n}G_{k+q}\left(\omega_n+\Omega_n\right)\chi_{q}\left(\Omega_n\right),
\end{equation}
where $G_k\left(\omega_n\right)$ is the electron propagator, $\chi_q\left(\omega_n\right)$ is the boson propagator, $h_k$ is a form factor for electron-boson coupling deduced from the symmetries of the system and  $k\gg q$. We have two regions to consider, the first being the case $q_z\gg q_{1,2}$. At a high enough frequency $\left|\omega_n\right|\gg T_{\ast}$, we can neglect the mass term and we have an electron-only theory with the usual self energy~\citep{Paul2017,garst2010electron,metzner2003soft} $\Sigma\left(i\omega_n\right)\propto \left|\omega\right|^{2/3}$ in 2D and $\Sigma\left(i\omega_n\right)\propto \omega_n\log\left|\omega_n\right|$ in 3D. Below this temperature the leading contribution to the self energy is  Fermi liquid-like.

At low frequency we then need to consider the contribution from $q\approx q_{1,2}$. We would expect hot spots in the regions where the critical directions are parallel to the Fermi surface, we find that there exists a contribution to the self energy for a damped system $\Sigma\left(i\omega_n\right)\propto i\omega\left(\left|\omega\right|/T_{\ast}\right)^{1/3}$. The form factor for the self-energy is given by
\begin{equation}
\begin{split}
[h_{\text{tot}}\left(k\right)]_{\alpha\beta}^2&=\left(\left(\cos\left(k_x\right)-\cos\left(k_y\right)\right)^2\right.\\
&\left.+\left(\sin\left(k_x\right)^2+\sin\left(k_y\right)^2\right)\sin\left(k_z\right)^2\right)\delta_{\alpha\beta},
\end{split}
\end{equation}
such that any terms on the off-diagonal cancel and we do not have a self-energy of matrix form. At the critical directions $h_k^2=0$ and so the hot spots are rendered cold.

\subsection{Elastocaloric effect}
%Inline with the nematic system, we can also calculate the elastocaloric effect in altermagnetic systems. A classical analysis yields
%\begin{equation}
%$\eta=\eta_{\text{bg}}-\lambda_H^2\left(\frac{T}{c_v}\right)\frac{d\chi}{dT}\epsilon=\eta_{\text{bg}}\lambda_H^2\left(\frac{T}{ac_v\left(T-%T_c\right)^2}\right)\epsilon,\label{eq::elastoclassic}
%\end{equation}
%in complete analogy with the nematic case~\citep{ikeda2021elastocaloric} and also found in Ref.~\cite{ye2024measurement}. We see that for an %altermagnet, it is is proportional to $H_z^2$, such that changing the magnetic field should invoke a changed response. For the case that %$\epsilon$  transforms according to $A_{1g}$, the elastocaloric effect is the same as the nematic case.

%We can also carry out the calculation in the quantum regime, where we must now include the effects of the Landau damping and renormalized mass %due to the magnetoelastic coupling. 

The analysis of the elastocaloric effect also proceeds along very similar steps as for nematic systems and yields
\begin{align}
    \eta &\approx -72 \Gamma_{*,0}T_* 
    \begin{cases}
        \frac{1}{9} \frac{\left(T/T_{\ast}\right)\log\frac{T_{\ast}}{T}}{\log\frac{T_{0}}{T_{\ast}}}, & T\ll T_{\ast}, \\[7pt]
        2\pi b \frac{\left(T/T_{*}\right)^{1/3}}{\log\frac{T_{0}}{T}}, & T\gg T_{\ast},
    \end{cases}
\end{align}
where $\Gamma_{*,0}$ is introduced in Eq.~\eqref{eq:Gammastar0}.

\subsection{Field-tuning of the elastic QCP in altermagnets} \label{sec::ExpAltermag}
In an altermagnetic system, the coupling strength can be tuned by an external magnetic field. This tunability provides a method to approach the elastic QCP from the Fermi liquid regime, characterised by a finite $m$ value, by increasing the magnetic field; see Fig.~\ref{fig::zeroT}. Changes of the Sommerfeld coefficient and elastocaloric effects due to variations in the magnetic field therefore offer an interesting  perspective to identify the elastic QCP .
To calculate the $\gamma$ and $\eta$ for a general value of magnetic field value, we use the following approximated renormalized mass obtained from Eq.~\eqref{eq::RenormalizedMassHzNeqZeroFull}: 
\begin{align}
    \frac{m(\mathbf{q},\mathbf{H})^2}{m_{\ast H}^2}\approx \left\{ \begin{array}{ll}1-\frac{H^2}{H_{\rm cr}^2}+\frac{H^2}{H_{\rm cr}^2}\frac{ q_2^2+ q_z^2}{q_1^2}, & |q_1|\gg |q_2|,|q_z|,\\
    1-\frac{H^2}{H_{\rm cr}^2}+\frac{H^2}{H_{\rm cr}^2}\frac{ q_1^2+ q_z^2}{q_2^2}, & |q_2|\gg |q_1|,|q_z|,\\
    1, & |q_z|\gg |q_1|,|q_2|,
    \end{array}\right.\label{eq::nonZeroHzRenormalizedMassWithGenH}
\end{align}
where $H_{\rm cr}=\sqrt{\frac{2(C_{11}-C_{12})m_{\ast H}^2}{\lambda^2}}$. 
When $H=0$, it reduces to $m_{\ast}$ in every momentum space which corresponds to the case of Fermi liquid regime, while it reduces to Eq.~\eqref{eq::nonZeroHzRenormalizedMass} when $H=H_{\rm cr}$ with $m_{\ast H}=m_{\ast}$

\subsubsection{Specific heat}
\begin{figure}
     \centering
     \includegraphics[width=0.48\textwidth]{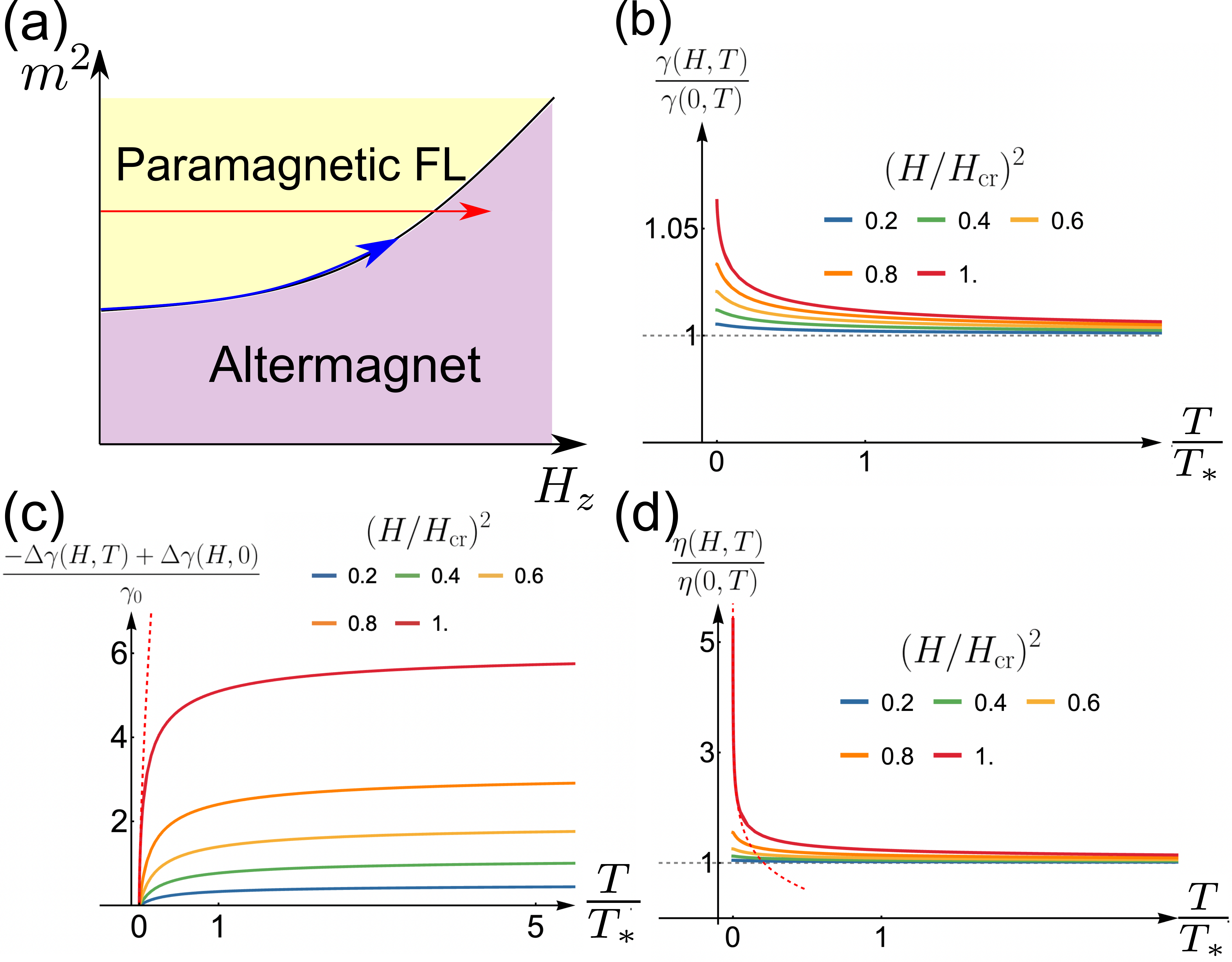}
     \caption{Constructing the phase diagram in the $T=0$ plane (a), makes clear the fact that an increased magnetic field creates an enhanced altermagnetic regime, allowing for a tunable QCP. In an experiment, we can measure the heat capacity along two lines, along the critical line (blue), or by using a magnetic field to tune through the critical point (red). The second option is the easiest to carry out in an experiment. We see that as the magnetic field reaches the critical value, the Sommerfeld coefficient (panel (b)) is not sensitive to the gapless modes and thus remains Fermi liquid-like, while the presence of the gapless modes can be more easily detected through measurements of $\Delta\gamma$ (panel (c)) where the dotted line shows the $T^{2/3}$ behaviour exhibited in the low temperature regime. The clearest indication of a critical point, however, comes from elastocaloric measurements (panel (d)) which diverge logarithmically (dotted line) in the low temperature regime.}
     \label{fig::zeroT}
 \end{figure}

In the Fermi liquid regime near the bare QCP (i.e., at zero magnetic field), the Sommerfeld coefficient shows the following behaviour~\cite{Paul2017},
\begin{align}
    \gamma(T)&\approx \frac{2\pi^3}{9}\gamma_0\begin{cases}
    \log \frac{T_0}{T_{\rm FL}}, & T\ll T_{\rm{FL}},\\[7pt]
    \log \frac{T_0}{T}, & T\gg T_{\rm{FL}},
    \end{cases}
    \label{eq::ZeroHSpecificHeat}
\end{align}
where the temperature $T_{\rm{FL}}(m)$ is the bare quantum critical crossover temperature between the paramagnetic Fermi liquid phase at low temperature and the bare quantum critical phase at high temperature as shown in Fig.~\ref{fig::phase_01} and Fig.~\ref{fig::3Dphase} and given by
\begin{equation}
    T_{\rm{FL}}(m)=\left(\frac{m}{\nu_\phi \Lambda}\right)^3 T_0.\label{eq:TFL}
\end{equation}
Thus the heat capacity shows qualitatively the same behaviour in the paramagnetic and elastic QC regimes, meaning that the heat capacity cannot distinguish between these two regimes. We see this in Fig.~\ref{fig::heatelasto} and this is also shown with several different magnetic fields in Fig.~\ref{fig::zeroT}(b).

To remedy this problem and see the effect of the gapless excitations induced by the magnetoelastic effect, we consider $\Delta \gamma(T,H)$, given by 
\begin{align}
    \Delta \gamma(T,H)\equiv \gamma(T,H)-\gamma(T,0),
\end{align}
for a fixed $m$. 

In $\Delta \gamma(T,H)$, the contributions coming from the phase space of $m(\mathbf{q})\approx m$ will be canceled, while those of the phase space of the reduced boson mass become more significant. To demonstrate this, we consider $\Delta \gamma(T,H)$ at the elastic quantum criticality ($\left.m\left(\mathbf{q}\right)\right|_{\rm min}=0$). We find that 
\begin{align}
  \Delta \gamma(T,H_{\rm{cr}})\approx \frac{4\pi}{3}\gamma_0\Big(c_0-c_1\left(\frac{T}{T_{\ast}}\right)^{2/3}\Big), & \;\; T\ll T_{\ast},\label{eq::DeltaGamma}
\end{align}
where $c_0$ and $c_1$ are constants independent of the momentum cutoff, magnetic field, temperature, and $m$.

 The $T^{2/3}$-dependent contribution has its origins in the gapless boson mode. This has also been discussed in Ref.~\citep{Paul2017}. However, it is not easy to separate the large contribution from the non-critical momentum space in the nematic case which makes an experimental observation difficult. Due to its tunability by the magnetic field in the altermagnetic case, it is more feasible to extract the $T^{2/3}$ behaviour originating from the gapless boson. Additionally, it is a general result regardless of the Fermi surface shape. 
 
Fig.~\ref{fig::zeroT}(c) shows $\Delta\gamma(T,H) - \Delta\gamma(0,H)$ for various magnetic fields. As the magnetic field approaches its critical value, corresponding to the location of the elastic QCP, it exhibits $T^{2/3}$ behaviour in the low-temperature limit, indicated by the red dashed line.

\subsubsection{Elastocaloric change}

Now let us consider the elastocaloric change. In the Fermi liquid regime, the bare mass $m^2$ of the altermagnetic boson can be tuned by the strain transforming according to the $A_{1g}$ irreducible representation. For $D_{4h}$, these strains correspond to $\epsilon_{xx}+\epsilon_{yy}$ and $\epsilon_{zz}$, such that $m^2(\epsilon)\sim c_{xx-yy}(\epsilon_{xx}+\epsilon_{yy})+c_{zz}\epsilon_{zz}$. According to Ref.~\citep{Zhu2003}, the elastocaloric change near the bare quantum critical point with the finite $m^2$ of an altermagnet is given by:
\begin{align}
    \eta (\epsilon,T,H=0) &\approx -72 \Gamma_{{\rm FL},0}T_{\rm FL}
    \begin{cases}
        \frac{1}{9} \frac{\left(T/T_{\rm FL}\right)}{\log\frac{T_{0}}{T_{\rm FL}}}, & T\ll T_{\rm FL}, \\[7pt]
        2\pi b \frac{\left(T/T_{\rm FL}\right)^{1/3}}{\log\frac{T_{0}}{T}}, & T\gg T_{\rm FL},
    \end{cases}
\end{align}
where $\Gamma_{{\rm FL},0}=\frac{\Lambda^3}{32\pi^4m^2}\frac{\partial m^2}{\partial \epsilon}\frac{1}{\gamma_0 T_0}$ and $T_{\rm{FL}}$ is given by Eq.~\eqref{eq:TFL} and now depends on the $A_{1g}$ strain $\epsilon$ through $m^2(\epsilon)$. If the bare mass $m$ is identical to $m_{*,H}$, values of $T_{\rm FL}$ and $\Gamma_{{\rm FL},0}$ are identical to values of $T_{*}$ and $\Gamma_{*,0}$ respectively. We see that this behaviour is qualitatively different to the results from the elastic QC regime in the low temperature regime.
More explicitly, $\frac{\eta(H)}{\eta(H=0)}$, where $H$ is the magnetic field, exhibits $\log T$ enhancement as the magnetic field approaches its critical value, corresponding to the location of the elastic QCP in the low-temperature regime, as shown in Fig.~\ref{fig::zeroT}(d).

Let us estimate the regime where the field-induced altermagnetic coupling
to the lattice becomes measurable. To estimate the magnitude of the piezomagnetic coupling we need
to determine the dimensionless ratio $\lambda\mu_{B}B/\left(C^{1/2}J\right)$
with $J=v_{\phi}\Lambda$ that determines the ratio $T_\ast/T_0$. We write $\lambda=\zeta/\sqrt{JV_{0}}$
with dimensionless strength $\zeta$ and $V_{0}$ the volume of the
unit cell. If $\Lambda\sim k_{F}$
and the dimensionless coupling constant $\alpha\sim1$, $T_{0}$ is
a significant fraction of $E_{F}$.  Suppose we want to achieve at least $T_{\ast}/T_{0}\sim10^{-3}$
to have an experimentally observable crossover regime.  At a magnetic field $B=10\,{\rm T}$ and
using $C=100\,{\rm GPa}$, $J/k_{{\rm B}}=100\,{\rm K}$, and $V_{0}=a^{3}$
with $a=5\times10^{-10}\:{\rm m}$  one gets $\zeta\sim10^{2}$. This
is significantly larger than values that have been measured in insulating
altermagnets, such as MnTe~\cite{aoyama2024piezomagnetic}, where $\zeta\sim0.1$. However, it is
sensible to expect piezomagnetic effects in metallic systems to be
significantly larger; in some cases giant piezomagnetism with values several order of magnitude have indeed been discussed~\cite{Ma2021multifunctional}. In particular for soft materials with a large $g$-factor
it is possible to reach values $T_{\ast}/T_{0}\sim10^{-3}$ with $\zeta\sim 1\sim  5$.
Hence, the observation of elastic quantum criticality in altermagnets, 
while likely more challenging than in nematic systems, is not out of
reach, where in iron-based superconductors such as $\text{Ba}\left(\text{Fe}_{1-x}\text{Co}_x\right)_2\text{As}_2$, a conservative estimate for the crossover temperature gives a value of at least \si{10}{K}~\cite{Paul2017}.

\section{Scaling of the elastocaloric effect}
\label{sec:scaling}
The scaling theory of quantum critical elasticity was formulated in
Ref.~\cite{Zacharias2015quantum} and is based upon the critical contribution to the free energy
density that obeys
\begin{equation}
f_{{\rm cr}}\left(r,T\right)=b^{-\left(d_{{\rm {\rm eff}}}+z\right)}f_{{\rm cr}}\left(b^{1/\nu}r,b^{z}T\right),\label{eq:scaling1}
\end{equation}
with correlation length exponent $\nu$, dynamic scaling exponent
$z$, and effective dimension $d_{{\rm eff}}$, while $r$ is a dimensionless measure of the distance to the QCP; in our case $r\propto m^2-m_\ast^2$. The emergence of an effective dimension is a consequence
of the fact that at criticality the effective phonon spectrum behaves
as 
\begin{equation}
\omega\left(\boldsymbol{q}\right)^{2}\sim c^{2}\boldsymbol{q}_{\perp}^{2}+aq_{{\rm soft}}^{4},\label{eq:disp_soft_hard}
\end{equation}
with $q_{{\rm soft}}$ parametrizing the soft line and $\boldsymbol{q}_{\perp}$
the $d-1$ components orthogonal to it. As discussed in Ref.~\cite{schwabl1985propagation,Zacharias2015quantum}, this
amounts to an effective dimension $d_{{\rm eff}}=2\left(d-1\right)+1$.
For $d=3$ it follows that $d_{{\rm eff}}=5$. In addition, for the over-damped
dynamics at a nematic or a field-tuned altermagnetic QCP $z=3$,
while the correlation-length exponent takes the mean-field value $\nu=\tfrac{1}{2}$.
If we make the choice $b=T^{-1/z}$ for the scaling factor in Eq.~\eqref{eq:scaling1},
we obtain for the critical contribution  $s_{{\rm cr}}=-\partial f_{{\rm cr}}/\partial T$ to the entropy density
\begin{equation}
s_{{\rm cr}}\left(r,T\right)=T^{\frac{d_{{\rm eff}}}{z}}\Phi\left(r T^{-\frac{1}{\nu z}}\right),
\end{equation}
with scaling function $\Phi\left(x\right)$. Thus at the critical
point using $r\propto\epsilon_0$ for symmetry-preserving strain, the temperature and strain derivatives equal:
\begin{eqnarray}
\frac{\partial s_{{\rm cr}}}{\partial T} & = & u_{T}T^{\frac{d_{{\rm eff}}-z}{z}},\nonumber \\
\frac{\partial s_{{\rm cr}}}{\partial\epsilon_0} & = & u_{\epsilon}T^{\frac{d_{{\rm eff}}-\nu^{-1}}{z}}.
\end{eqnarray}
$u_{T}$ and $u_{\epsilon}$ are constant coefficients. If we now add those critical contributions to the Fermi liquid background terms we obtain for the elastocaloric effect divided by $T$ (which is the appropriate dimensionless quantity that scales like the Gr\"uneisen parameter):
\begin{eqnarray}
\eta/T & = & -\frac{\frac{\partial s_{{\rm FL}}}{\partial\epsilon_0}+\frac{\partial s_{{\rm cr}}}{\partial\epsilon_0}}{T\frac{\partial s_{{\rm FL}}}{\partial T}+T\frac{\partial s_{{\rm cr}}}{\partial T}}\nonumber \\
 & = & \frac{-\Gamma_{\rm FL}\gamma_{{\rm FL}}-u_{\epsilon}T^{\frac{d_{{\rm eff}}-\nu^{-1}-z}{z}}}{\gamma_{{\rm FL}}+u_{T}T^{\frac{d_{{\rm eff}}-z}{z}}},
\end{eqnarray}
where $\Gamma_{\rm FL}$ is the constant Gr\"uneisen parameter of a Fermi liquid. Let us first analyse the denominator, i.e.\ $c/T$ for $d_{{\rm eff}}=5$,
$\nu=1/2$, and $z=3$.
We see that $c/T=\gamma_{{\rm FL}}+u_{T}T^{2/3}\approx\gamma_{{\rm FL}}$,
i.e., the critical contribution to the heat capacity is sub-leading.
The situation is different for the numerator, where the exponent $\frac{d_{{\rm eff}}-\nu^{-1}-z}{z}\rightarrow0$,
which should be interpreted as logarithmic behaviour. Hence, for
the numerator, i.e.\ for the thermal expansion of the system, the critical
contribution is the dominant one such that 
\begin{equation}
\eta/T\propto\log T,
\end{equation}
fully consistent with our explicit analysis. Thus elastic quantum
criticality gives the dominant contribution to the thermal expansion and the elastocaloric
effect, but is sub-leading for the heat capacity.

The physical significance of our result lies in the fact that changes in strain or stress alter the excitation spectrum more dramatically than variations
in temperature. This is reflected in the larger scaling dimension associated with a strain derivative compared to that of a temperature variation. Consequently, quantities such as the elastocaloric effect, thermal expansion, and the Gr\"uneisen parameter exhibit a more singular response than the heat capacity or the single-particle excitation spectrum.

Our analysis is based upon a one-loop approach of the coupled electron-boson
problem, i.e. we closely follow the spirit of the Hertz-Millis~\cite{Hertz1976quantum,millis1993effect} approach of itinerant quantum critical points. As we include
the coupling to elastic modes, we stay consistently in three spatial dimensions.
In Ref.~\cite{abanov2004anomalous} it was found that the Hertz-Millis theory is
incomplete for $d+z\leq4$, i.e. there are issues with the theory
for systems with $d=z=2$. However, no inconsistencies were identified
in $d=3$. Singular behavior at higher loops was also analyzed for
$d=2$ in Refs~\cite{lee2009low,metlitski2010quantum1,metlitski2010quantum2}, yet no such behavior
was found in $d=3$. Finally, the theory discussed here, but without
strain coupling was analyzed in Ref.~\cite{chubukov2005superconductivity}, including an analysis
of vertex corrections, with the result that there are no singular corrections
occurring due to such vertex corrections. Hence, our approach
can be used to reliably analyze these three-dimensional QCP's, in particular
as the boson-mediated interaction in strain-coupled systems is less
singular than in the case without inclusion of strain modes. More concretely, in the quantum regime, a combination of an effective spatial dimension of $5$ and dynamical exponent $3$ means that a perturbative one-loop analysis is justified. With strain coupling, we would expect well-defined behavior  even if the actual spatial dimension were $2$.

\section{Summary}\label{sec::summary}
This paper explores the role of elastic fluctuations near electronic quantum critical points in nematic and altermagnetic systems that break rotational symmetries. It is well-established from analyses of classical and quantum phase fluctuations that coupling to elastic modes alters the universality class of a critical point in such systems. The long-range coupling mediated by acoustic phonons suppresses critical order parameter fluctuations. This suppression stems from the selective lattice softening of phonon modes at nematic or altermagnetic phase transitions, where the sound velocity vanishes only along isolated lines in momentum space. While these results were obtained for a coupling to $B_{1g}$ strain, they are valid generally for any nematic or altermagnetic order parameter which couples to symmetry breaking strain. In all cases a momentum dependent coupling emerges, leading to a direction dependent mass with minima along certain $q$ directions. Different irreducible representations have different form factors which lead to different directions along which the system becomes critical but this will not affect the behaviour of the elastocaloric effect. In cubic systems we note that $A_{1g}$ does not have a corresponding piezomagnetic response. For systems with a higher-dimensional irreducible representation, a cubic term emerges in the field theory (for altermagnetic systems this is field dependent), inducing a weak first-order transition and our theory no longer applies~\cite{Cowley1976}.

The main findings of this work are twofold. First, we show that elastic quantum criticality at a nematic quantum critical point is characterised by an elasto-caloric anomaly, leading to results inconsistent with a Fermi liquid.  This is in contrast to observables such as the specific heat capacity and the single-particle self-energy that exhibit Fermi liquid-like behaviour, in agreement with earlier results in Ref.~\cite{Zacharias2015quantum}.
Second, we show that similar behaviour arises in altermagnets subjected to an external magnetic field. The field-induced piezomagnetic coupling leads to effects analogous to those observed in nematic systems. Consequently, the elastocaloric effect could serve as a  tool to analyse  elastic quantum criticality in these systems as well. The crossover temperature between bare criticality -- identical to that without lattice coupling -- and elastic quantum criticality becomes a field-dependent scale, offering experimental tunability. To make our analysis of altermagnets more accessible, we provide a classification of symmetry-allowed couplings for various crystal symmetries.

The phenomena discussed here are particularly relevant to iron-based superconductors, where nematic quantum criticality has been studied extensively~\cite{Yoshizawa2012,Meingast2012,Boehmer2014,Kuo2016,Ikeda2021}. In these systems, the crossover to the elastic critical regime roughly coincides with the temperature range where elastic constants are significantly suppressed, making  the elastic-quantum critical regime clearly accessible. In altermagnets, however, the crossover temperature to elastic criticality is less well-established. It depends on both the strength of the piezomagnetic coupling and the magnetic field energy. While piezomagnetic coupling tends to be larger in itinerant systems, the relatively small energy scales of magnetic fields suggest that substantial crossover temperatures are more likely in systems with low overall altermagnetic order energies.
As the exploration of altermagnetic materials is still in its infancy, our theory may encourage the search for systems with strong piezomagnetic couplings, large magnetic $g$-factors, and moderate magnetic interaction scales. Overall, our results highlight quantum criticality in nematic and altermagnetic systems as an intriguing manifestation of the intricate coupling between electronic and lattice degrees of freedom.

\begin{acknowledgments}
We are grateful to Anzumaan Chakraborty, Rafael M. Fernandes, Ian R. Fisher, Elena Gati, Tobias Holder, Andrew P. Mackenzie, Rahel Ohlendorf, Indranil Paul, Jairo Sinova, Libor Šmejkal, and Roser Valentí for helpful discussions. This work was supported
by the German Research Foundation TRR 288-422213477 ELASTO-Q-MAT,
Projects A11 (M.G.) and A07 (G.P., I.J., J.S.), and the
DFG project SCHM 1031/12-1 (C.R.W.S.). 
\end{acknowledgments}

\bibliography{FTC.bib}

\appendix

\begin{widetext}

\section{Renormalized boson propagator for generic elastic constants}\label{App:RenormalizedMassGeneral}

Here we consider the renormalized mass of the boson due to the elastic coupling. We use the eigenmodes of the elasticity without the assumptions used in the main text and show that the results do not change qualitatively compared to those obtained with the aforementioned assumptions.

To get the eigenmodes of the elastic mode, we use the following dynamic matrix $M(\mathbf{q})$~\citep{Karahasanovic2016}.
\begin{align}
M(q)=\left(\begin{array}{ccc}c_{11}q_x^2+c_{66}q_y^2+ c_{44}q_z^2 & (c_{12}+c_{66})q_xq_y & (c_{13}+c_{44})q_xq_z \\ 
(c_{12}+c_{66})q_xq_y & c_{66}q_x^2+c_{11}q_y^2+c_{44}q_z^2 & (c_{13}+c_{44})q_yq_z \\ 
(c_{13}+c_{44})q_xq_z & (c_{13}+c_{44})q_yq_z & c_{44}(q_x^2+q_y^2)+c_{33}q_z^2
\end{array}\right).
\end{align}

\subsection{Numerical Results}
Before considering the analytical approach, we first obtain the numerical results for the direction selective mass when $\mathbf{H}=H\hat{z}$, which gives an identical result to the nematic case, and $\mathbf{H}=H\hat{x}$ using the values of $c_{ij}$ of SrRuO$_4$, which is a representative and well-studied system  which has a $D_{4h}$ symmetry~\citep{Ghosh2021}. 
\begin{align}
c_{11}\approx240 \si{GPa},\; c_{12}\approx140 \si{GPa},\; c_{13}\approx85 \si{GPa},\; c_{33}\approx257 \si{GPa},\; c_{44}\approx70 \si{GPa},\; c_{66}\approx65 \si{GPa}.\label{eq::CijSrRuO4}
\end{align}

\begin{figure}
\centering
\includegraphics[width=0.7\textwidth]{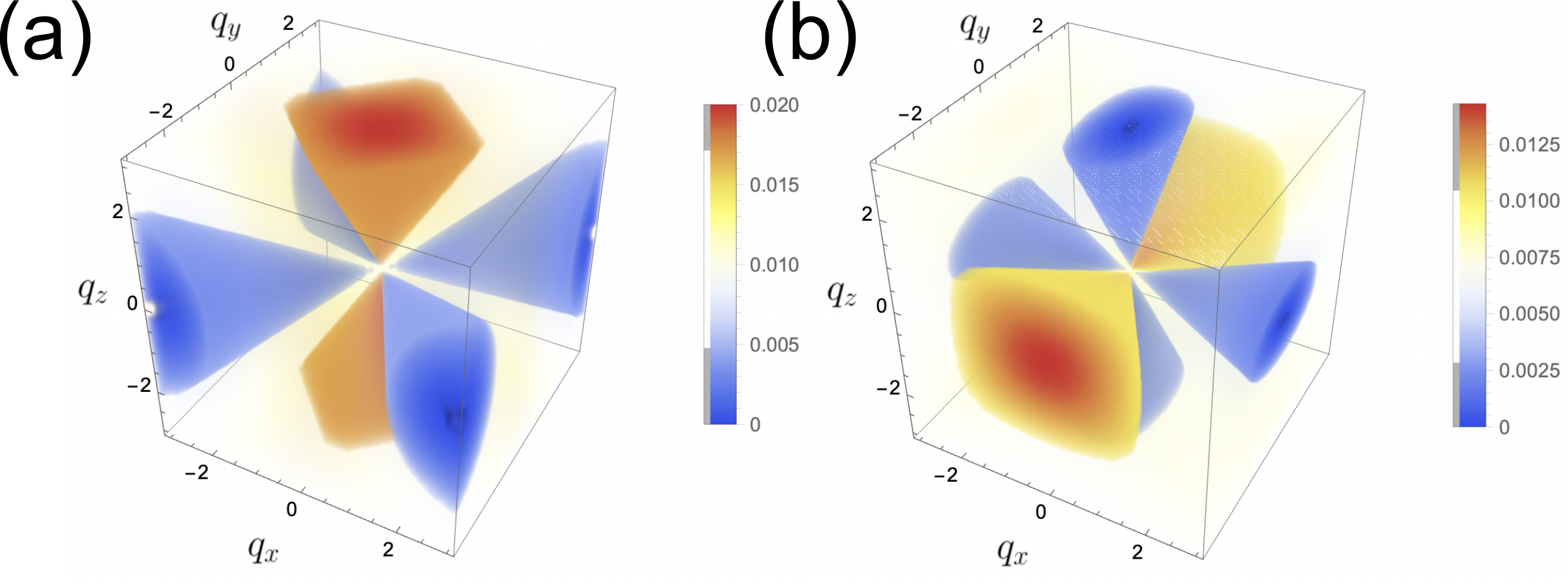}
\caption{Density plot of $m^2$ for (a) $\mathbf{H}=H\hat{z}$ and (b) $\mathbf{H}=H\hat{x}$ with values of $c_{ij}$ in Eq.~\eqref{eq::CijSrRuO4}. Soft lines in momentum space  and their vicinity and indicated in blue, while red marks regions with a particularly hard gap.}\label{fig::DirectionSelectiveMass_General_HzNeqZero}
\end{figure}

Fig.~\ref{fig::DirectionSelectiveMass_General_HzNeqZero} shows a similar direction-selective mass to that considered in the main text.

\subsection{Analytical results}
It is difficult to get full analytical results for general values of momentum and $c_{ij}$. However we do not need analytical results for the full momentum space, since we are going to consider only the momentum regimes which give the dominant contributions to the free energy (as we did in the main text). From the numerical results, we already know that the results do not change qualitatively. As such, we will use the same momentum regimes used in the main text. 

For case (a) $\mathbf{H}=H\hat{z}$, these regimes are given by $(\it{i})\; |q_1|\gg |q_2|,|q_z|$, $(\it{ii})\; |q_2|\gg |q_1|,|q_z|$, and $(\it{iii})\; |q_z|\gg |q_1|,|q_2|$ where $q_1=(q_x+q_y)/\sqrt{2}$ and $q_2=(q_x-q_y)/\sqrt{2}$. For case (b) $\mathbf{H}=H\hat{x}$, these regimes are given by $(\it{i})\; |q_z|\gg |q_x|,|q_y|$, $(\it{ii})\; |q_x|\gg |q_y|,|q_z|$, and $(\it{iii})\; |q_y|\gg |q_z|,|q_x|$.

Since there is a dominant component of the momentum in each regime, we can use a perturbation theory to obtain the eigenmodes of the dynamic matrix $M(\mathbf{q})$ by considering the matrix elements with the small momentum as a perturbation.

\subsubsection{$\mathbf{H}=H\hat{z}$ case (Identical to the Nematic case)}
We consider the three momentum regimes: ${\it (i)}$ $|q_1|\gg |q_2|,|q_z|$, ${\it (ii)}$ $|q_2|\gg |q_1|,|q_z|$ and ${\it (iii)}$ $|q_z|\gg |q_1|,|q_2|$. Then
\begin{align}
m^2(\mathbf{q},\mathbf{H})\approx \begin{cases}
    m^2_{\rm{min}} + (\lambda_zH_z)^2\frac{1}{(c_{11}-c_{12})^2(c_{11}+c_{12}+2c_{16})}\frac{4(c_{11}+c_{12})c_{66}q_2^2+c_{44}(c_{11}+c_{12}+2c_{66})q_z^2}{q_1^2}, &(\it{i})\; |q_1|\gg |q_2|,|q_z| ,\\
m^2_{\rm{min}} + (\lambda_zH_z)^2\frac{1}{(c_{11}-c_{12})^2(c_{11}+c_{12}+2c_{16})}\frac{4(c_{11}+c_{12})c_{66}q_1^2+c_{44}(c_{11}+c_{12}+2c_{66})q_z^2}{q_2^2}, &(\it{ii})\; |q_2|\gg |q_1|,|q_z|,\\
 m^2-\frac{(\lambda_\perp H_x)^2}{4}\frac{q_1^2+q_2^2}{c_{44}q_z^2}\approx m^2 ,& (\it{iii})\; |q_z|\gg |q_1|,|q_2|,
\end{cases}
\end{align}
where $m_{\rm{min}}^2=m^2-\frac{(\lambda_z H_z)^2}{2}\frac{1}{c_{11}-c_{12}}$. Replacing $\lambda_zH_z$ with $\lambda$ gives the renormalized mass of the nematic case. 

Note that $m^2_{\rm{min}}$ is exactly the same as Eq.~\eqref{eq::rminHzNeqZero} of the main text. At criticality $m_{\rm{min}}=0$ and so this result is qualitatively the same as the results obtained in the main text. 

\subsubsection{$\mathbf{H}=H\hat{x}$ case}

 We consider the three momentum regimes: ${\it (i)}$ $|q_x|\gg |q_y|,|q_z|$, ${\it (ii)}$ $|q_z|\gg|q_x|,\; q_y=0$ and ${\it (iii)}$ $|q_y|\gg|q_x|, \; q_z=0$. Then
\begin{align}
m^2(\mathbf{q},\mathbf{H})\approx \begin{cases}
    m^2_{\rm{min}}+\frac{(\lambda_\perp H_x)^2}{4c_{44}}\frac{c_{11}c_{44}q_x^2+(c_{11}c_{33}-c_{13}^2)q_z^2}{c_{11}c_{44} q_x^2}, & (\it{i})\; |q_x|\gg |q_y|,|q_z|,\\
    m^2_{\rm{min}}+\frac{ (\lambda_\perp H_x)^2}{4c_{44}}\frac{c_{11}c_{33}-c_{44}^2}{c_{33}c_{44}}q_x^2, & (\it{ii})\; |q_z|\gg|q_x|,\; q_y=0,\\
    m^2, &(\it{iii})\;  |q_y|\gg|q_x|, \; q_z=0,
\end{cases}
\end{align}
where $m^2_{\rm{min}}=m^2-\frac{  (H_x\lambda_\perp)^2 }{4 c_{44}}$.

Here we investigate the more restricted momentum regimes specified by $q_y=0$ and $q_z=0$ for cases $\it (ii)$ and $\it (iii)$ since we can still use non-degenerate perturbation theory. However, we expect qualitatively the same behaviour to $r(\mathbf{q},\mathbf{H})$ in Eq.~\eqref{eq::nonZeroHxRenormalizedMass} for the momentum spaces $|q_z|\gg|q_x|,|q_z|$ and $|q_y|\gg |q_x|,|q_z|$. This has been verified numerically.

Considering the critical case where $r_{\rm{min}}=0$, the renormalized mass is
\begin{align}
    m^2(\mathbf{q},\mathbf{H})\approx \left\{ \begin{array}{ll}
    m_{\ast H}^2\frac{q_y^2+q_z^2}{q_x^2},& (\it{i})\; |q_x|\gg|q_y|,|q_z|\\
m_{\ast H}^2\frac{q_x^2+q_y^2}{q_z^2}, & (\it{ii})\; |q_z|\gg |q_x|,|q_y|\\
m_{\ast H}^2, & (\it{iii})\; |q_y|\gg |q_x|,|q_z|
\end{array}\right.\label{eq::nonZeroHxRenormalizedMass}
\end{align}
where $m_{\ast H}^2=\frac{\lambda_\perp^2 H_x^2}{4c_{44}}$ and we set the coefficients to be 1 for simplicity.

\section{Specific heat near a nematic QCP}\label{app::specific_heat}
Here we provide full details of the calculation of specific heat performed  in Ref.~\cite{Paul2017}. 

The free energy density of the system is given by
\begin{align}
    &F=\frac{T}{2}\sum_{\omega}\int\frac{d^3q}{(2\pi)^3}\log\chi^{-1}(\omega,\mathbf{q})\nonumber \\
    &= -\int\frac{d^3q}{(2\pi)^3}\int_0^\infty \frac{d\omega }{2\pi}\coth\frac{\beta \omega}{2} \tan^{-1}\frac{\tfrac{\Xi^2 \omega}{|\mathbf{q}|v_{\rm{F}}}}{v_{\phi}^2|\mathbf{q}|^2+m^2(\mathbf{q})},
\end{align}
where the Matsubara frequency summation is evaluated via a contour integral. 
The Sommerfeld coefficient $\gamma(T)=-\frac{\partial F^2}{\partial^2T}$ is then given by 
\begin{align}
\gamma & =  \gamma_0
\int_{0}^{\infty}\frac{xdx}{e^{x}-1} \int_{0}^{\bar{\Lambda}}d^3\bar{q}\frac{|\bar{q}|^3\left(\bar{m}^{2}\left(\boldsymbol{q}\right)+\bar{q}^{2}\right)^{3}}{\left(\bar{q}^{2}\left(\bar{m}^{2}\left(\boldsymbol{q}\right)+\bar{q}^{2}\right)^{2}+x^{2}\bar{T}^2\right)^{2}},
   \label{eq:gamma_nem1}
\end{align}
where 
\begin{equation}
    \gamma_0=\frac{3}{\pi}\alpha\rho_{F}. 
\end{equation}
Here we used the following dimensionless variables
\begin{gather}
\bar{m}\left(\boldsymbol{q}\right)=\frac{m\left(\boldsymbol{q}\right)}{m_{\ast}},\quad\boldsymbol{\bar{q}}=\frac{v_{\phi}\boldsymbol{q}}{m_{\ast}},
    \quad \bar{T}=\frac{T}{T_{\ast}}
\end{gather}
as well as $\bar{\Lambda}=\left(T_{0}/T_{\ast}\right)^{1/3}$, and the crossover temperature scale
\begin{equation}
 T_{\ast}=\left(\frac{m_{\ast}}{v_{\phi}\Lambda}\right)^{3}T_{0}.
\end{equation}
  $T_\ast$ vanishes in the limit of zero nemato-elastic coupling $\lambda$ and the analysis of the integral  is straightforward, giving the logarithmic dependence of the Sommerfeld coefficient of Eq.~\eqref{eq:gamma_0}.

For temperatures $T \gg T_*$ the result remains unaffected by the nemato-elastic coupling and Eq.~\eqref{eq:gamma_0} still holds. In the other limit $T \ll T_*$ the specific heat coefficient simplifies to 
\begin{align}
\gamma &\approx \gamma_0 \frac{\pi^2}{6} \int_{0}^{\bar{\Lambda}}d^3\bar{q}
\frac{1}{\bar q (\bar{m}^{2}\left(\boldsymbol{q}\right)+\bar{q}^{2})}.
\end{align}
Using that $\bar m(\boldsymbol{q}) = \bar m(\hat q)$ depends only on the orientation of momentum $\hat q = \boldsymbol{q}/q$, but not on the amplitude $q = |\boldsymbol{q}|$, this reduces to 
\begin{align}
\gamma &\approx \gamma_0 \frac{2 \pi^3}{3} 
\int \frac{d \hat q}{4\pi} \log \frac{\bar \Lambda}{\bar m(\hat q)} \approx \gamma_0 \frac{2 \pi^3}{9} \log \frac{T_0}{T_*}, 
\end{align}
where the last approximation holds for exponentially large $T_0/T_*$. Importantly, the integral over the orientation of $\boldsymbol{q}$, $\int d\hat q \log \bar m(\hat q)$ is well-behaved and convergent. Potentially dangerous are the soft directions of $\boldsymbol{q}$, i.e., regimes $(i)$ and $(ii)$ in Eq.~\eqref{eq:rapprox}. These two regimes are equivalent by symmetry, so it suffices to analyse only one of them, e.g.~regime $(i)$. Parametrizing the wavevector with spherical coordinates 
\begin{equation}
    \bar{q}_1=\bar{q}\cos\theta,\; \bar{q}_2=\bar{q}\sin\theta\cos\phi,\;
    \bar{q}_z=\bar{q}\sin\theta\sin\phi ,
\end{equation}
regime $(i)$ then corresponds to small values of the polar angle, say $\theta \in [0,\theta_0]$, where $\theta_0$ is a delimiting value. The contribution to the angular integral is then of the form $\int_0^{\theta_0} d\theta \theta \log \theta$ which is indeed convergent for small $\theta$.

In summary, for the specific heat coefficient we thus obtain
\begin{align}
    \gamma(T) \approx \frac{2\pi^{3}}{9}\gamma_0 \begin{cases}
    \log \frac{T_0}{T_\ast}, & T\ll T_{\ast},\\
    \log \frac{T_0}{T}, & T\gg T_{\ast}.
    \end{cases}
\end{align}
Below the crossover temperature $T_\ast $ the specific heat is Fermi liquid-like, while above $T_\ast $  the system is in the bare QC regime and behaves like a marginal Fermi liquid,  as reported in Ref.~\cite{Paul2017}.
\color{black}

%\section{Field-induced lattice softening and piezomagnetic coupling in altermagnets}\label{sec::softening}

\section{Piezomagnetic coupling for selected point groups}\label{sec::softening}

In this appendix, we summarize the symmetry-allowed piezomagnetic couplings for a selected number of point groups.
The table that describes tetragonal systems with point group $D_{4h}$ was given in Table~\ref{Tab1} of the main text.
Here we list the allowed piezomagnetic coupling terms for orthorhombic systems with point group $D_{2h}$ in Table~\ref{Tab4}, for hexagonal systems with point group $D_{6h}$ in Table~\ref{Tab2}, and for cubic systems with point group $O_h$ in Table~\ref{Tab3}.
The irreducible representations that do not allow for a ferromagnetic moment are indicated in solid black, while the representations that will, inevitably, be accompanied by a ferromagnetic component of the magnetization are marked in gray.
Piezomagnetic coupling occurs for all symmetries in all point groups, with the single exception of the irrep $A_{1g}^-$ of the cubic group $O_h$.
Here a non-linear strain coupling, quadratic in the strain-tensor elements, is needed to couple a magnetic field to the altermagnetic order parameter.

\begin{table}[b]
{\renewcommand{\arraystretch}{1.3}
\renewcommand{\tabcolsep}{5.2pt}
\begin{tabular}{c|c|c}
\hline \hline
\multicolumn{3}{c}{\normalsize $D_{2h}$ ($mmm$) point group} \tabularnewline
AM irrep & coupling to fermions & piezomagnetic coupling \tabularnewline
\hline
$A_{1g}^{-}$ & \begin{tabular}{c}
$g \phi k_y k_z \sigma_x$\tabularnewline
$g' \phi k_x k_z \sigma_y$\tabularnewline
$g'' \phi k_x k_y \sigma_z$\tabularnewline
\end{tabular} & \begin{tabular}{c}
$\lambda\phi\epsilon_{yz}H_x$\tabularnewline
$\lambda'\phi\epsilon_{xz}H_y$\tabularnewline
$\lambda''\phi\epsilon_{xy}H_z$\tabularnewline
\end{tabular} \tabularnewline

\hline
FM irrep & & \tabularnewline
\hline
\textcolor{FMirrepColour}{$B_{1g}^{-}$} & \textcolor{FMirrepColour}{$g \phi \sigma_z$} & \begin{tabular}{c}
\textcolor{FMirrepColour}{$\lambda\phi\epsilon_{A_{1g}}H_z$}\tabularnewline
\textcolor{FMirrepColour}{$\lambda'\phi\epsilon_{xz}H_x$}\tabularnewline
\textcolor{FMirrepColour}{$\lambda''\phi\epsilon_{yz}H_y$}\tabularnewline
\end{tabular} \tabularnewline
\hline
\textcolor{FMirrepColour}{$B_{2g}^{-}$} & \textcolor{FMirrepColour}{$g \phi \sigma_y$} & \begin{tabular}{c}
\textcolor{FMirrepColour}{$\lambda\phi\epsilon_{A_{1g}}H_y$}\tabularnewline
\textcolor{FMirrepColour}{$\lambda'\phi\epsilon_{xy}H_x$}\tabularnewline
\textcolor{FMirrepColour}{$\lambda''\phi\epsilon_{yz}H_z$}\tabularnewline
\end{tabular} \tabularnewline
\hline 
\textcolor{FMirrepColour}{$B_{3g}^{-}$} & \textcolor{FMirrepColour}{$g \phi \sigma_x$} & \begin{tabular}{c}
\textcolor{FMirrepColour}{$\lambda\phi\epsilon_{A_{1g}}H_x$}\tabularnewline
\textcolor{FMirrepColour}{$\lambda'\phi\epsilon_{xy}H_y$}\tabularnewline
\textcolor{FMirrepColour}{$\lambda''\phi\epsilon_{xz}H_z$}\tabularnewline
\end{tabular} \tabularnewline
\hline \hline
\end{tabular}}
\caption{The coupling of altermagnetic (AM) and ferromagnetic (FM) order parameters $\phi$ to fermions and to simultaneous magnetic and strain fields (piezomagnetism) for the orthorhombic point group $D_{2h}$ ($mmm$).
The first column indicates the irrep of $\phi$.
$g$ and $\lambda$ are coupling constants, $k_{\alpha}$ is the momentum, $\sigma_{\alpha}$ are Pauli matrices, $\epsilon_{\alpha\beta}$ is the strain tensor, and $H_{\alpha}$ is the magnetic field.
The possibility of having $g' \neq g$ and $\lambda' \neq \lambda$ reflects magnetic anisotropy.
$\epsilon_{A_{1g}}$ are the trivially transforming strain components.
The coupling to lattice fermions is deduced by replacing $k_{\alpha} k_{\beta}$ with $\sin k_{\alpha} \sin k_{\beta}$.
In contrast to the tetragonal case (Table~\ref{Tab1}), now there is only one altermagnetic state which is a combination of a magnetic dipole with a charge quadrupole.
All other representations correspond to ferromagnetic orders along the crystalline axes, but display the same piezomagnetic behaviour for field directions orthogonal to the easy axis.}
\label{Tab4}
\end{table}

\begin{table*}
{\renewcommand{\arraystretch}{1.3}
\renewcommand{\tabcolsep}{7.6pt}
\begin{tabular}{c|c|c}
\hline \hline
\multicolumn{3}{c}{\normalsize $D_{6h}$ ($6/mmm$) point group} \tabularnewline
AM irrep & coupling to fermions & piezomagnetic coupling \tabularnewline
\hline
$A_{1g}^{-}$ & $g \phi k_z \! \left(k_y \sigma_x - k_x \sigma_y\right)$ & $\lambda\phi\left(\epsilon_{yz} H_x - \epsilon_{xz} H_y\right)$ \tabularnewline
\hline 
$B_{1g}^{-}$ & $g \phi\left(\left(k_x^2 - k_y^2\right) \sigma_x - 2 k_x k_y \sigma_y\right)$ & $\lambda\phi\left(\epsilon_{x^2-y^2} H_x - 2\epsilon_{xy} H_y\right)$ \tabularnewline
\hline 
$B_{2g}^{-}$ & $g \phi\left(2 k_x k_y \sigma_x + \left(k_x^2 - k_y^2\right) \sigma_y\right)$ & $\lambda\phi\left(2 \epsilon_{xy} H_x + \epsilon_{x^2-y^2} H_y\right)$ \tabularnewline
\hline 
$E_{2g}^{-}$ & \begin{tabular}{c}
$g k_z \! \left(\phi_1 \left(k_y \sigma_x + k_x \sigma_y\right) - \phi_2 \left(k_x \sigma_x - k_y \sigma_y\right)\right)$ \tabularnewline
$g' \left(2 \phi_1 k_x k_y - \phi_2 \left(k_x^2 - k_y^2\right)\right) \sigma_z$
\end{tabular} & \begin{tabular}{c}
$\lambda \left(\phi_1\left(\epsilon_{yz} H_x + \epsilon_{xz} H_y\right) - \phi_2 \left(\epsilon_{xz} H_x - \epsilon_{yz} H_y\right)\right)$ \tabularnewline
$\lambda' \left(2 \phi_1 \epsilon_{xy} - \phi_2 \epsilon_{x^2-y^2} \right) H_z$
\end{tabular} \tabularnewline

\hline
FM irrep & & \tabularnewline
\hline
${\color{FMirrepColour}A_{2g}^{-}}$ & \textcolor{FMirrepColour}{$g \phi \sigma_z$} & \begin{tabular}{c}
\textcolor{FMirrepColour}{$\lambda\phi\epsilon_{A_{1g}}H_z$}\tabularnewline
\textcolor{FMirrepColour}{$\lambda'\phi\left(\epsilon_{xz} H_x + \epsilon_{yz} H_y\right)$}
\end{tabular} \tabularnewline
\hline 
\textcolor{FMirrepColour}{$E_{1g}^{-}$} & \textcolor{FMirrepColour}{$g \left(\phi_1 \sigma_x + \phi_2 \sigma_y\right)$} & \begin{tabular}{c}
\textcolor{FMirrepColour}{$\lambda\epsilon_{A_{1g}}\left(\phi_1 H_x + \phi_2 H_y\right)$} \tabularnewline
\textcolor{FMirrepColour}{$\lambda' \left(\phi_1\left(\epsilon_{x^2-y^2} H_x + 2 \epsilon_{xy} H_y\right) + \phi_2 \left(2 \epsilon_{xy} H_x - \epsilon_{x^2-y^2} H_y\right)\right)$} \tabularnewline
\textcolor{FMirrepColour}{$\lambda'' \left(\phi_1 \epsilon_{xz} + \phi_2 \epsilon_{yz}\right) H_z$}
\end{tabular} \tabularnewline
\hline \hline
\end{tabular}}
\caption{The coupling of altermagnetic (AM) and ferromagnetic (FM) order parameters $\phi$ to fermions and to simultaneous magnetic and strain fields (piezomagnetism) for the hexagonal point group $D_{6h}$ ($6/mmm$).
Note that the hexagonal $a$ axis is parallel to the $y$ axis~\cite[Table 10.1]{Dresselhaus2007} so $x^3 - 3 x y^2 \in B_{1u}$ and $3 x^2 y - y^3 \in B_{2u}$.
The first column indicates the irrep of $\phi$.
For $E_{2g}$, $(\phi_1,\phi_2)$ transforms like $(x^2-y^2,2xy)$, while for $E_{1g}$ $(\phi_1,\phi_2)$ transforms like $(\sigma_x,\sigma_y)$.
$g$ and $\lambda$ are coupling constants, $k_{\alpha}$ is the momentum, $\sigma_{\alpha}$ are Pauli matrices, $\epsilon_{\alpha\beta}$ is the strain tensor, and $H_{\alpha}$ is the magnetic field.
The possibility of having $g' \neq g$ and $\lambda' \neq \lambda$ reflects magnetic anisotropy.
$\epsilon_{A_{1g}}$ are the trivially transforming strain components.}
\label{Tab2}
\end{table*}

\begin{table*}
{\renewcommand{\arraystretch}{1.4}
\renewcommand{\tabcolsep}{4.0pt}
\begin{tabular}{c|c|c}
\hline \hline
\multicolumn{3}{c}{\normalsize $O_h$ ($m{-}3m$) point group} \tabularnewline
AM irrep & coupling to fermions & piezomagnetic coupling \tabularnewline
\hline
$A_{1g}^{-}$ & \begin{tabular}{c}
$g \phi \big(k_y k_z (k_y^2 - k_z^2) \sigma_x - k_x k_z (k_x^2 - k_z^2) \sigma_y$ \tabularnewline
$+ \, k_x k_y (k_x^2 - k_y^2) \sigma_z\big)$
\end{tabular} &
$\lambda \phi \left(\epsilon_{yz} \epsilon_{y^2-z^2} H_x - \epsilon_{xz} \epsilon_{x^2-y^2} H_y + \epsilon_{xy} \epsilon_{x^2-y^2} H_z\right)$ \tabularnewline
\hline 
$A_{2g}^{-}$ & $g \phi \left(k_y k_z \sigma_x + k_x k_z \sigma_y + k_x k_y \sigma_z\right)$ & $\lambda\phi\left(\epsilon_{yz} H_x + \epsilon_{xz} H_y + \epsilon_{xy} H_z\right)$ \tabularnewline
\hline
& & \\[-12pt]
$E_{g}^{-}$ & $\begin{aligned}
g \big(&\phi_1 \sqrt{3} k_z \! \left(k_y \sigma_x - k_x \sigma_y\right) \\
&- \phi_2 \left(k_y k_z \sigma_x + k_y k_z \sigma_y - 2 k_x k_y \sigma_z\right)\big)
\end{aligned}$ & $\lambda \left(\phi_1 \sqrt{3} \left(\epsilon_{yz} H_x - \epsilon_{xz} H_y\right) - \phi_2\left(\epsilon_{yz} H_x + \epsilon_{xz} H_y - 2 \epsilon_{xy} H_z\right)\right)$ \\[-12pt]
& & \tabularnewline
\hline 
& & \\[-12pt]
$T_{2g}^{-}$ & \begin{tabular}{c}
$\begin{aligned}
g &\Big(\phi_1 k_x \! \left(k_y \sigma_y - k_z \sigma_z\right) \\
&\quad - \phi_2 k_y \! \left(k_x \sigma_x - k_z \sigma_z\right) \\
&\qquad + \phi_3 k_z \! \left(k_x \sigma_x - k_y \sigma_y\right)\Big)
\end{aligned}$ \\[22pt]
$g' \bigg(\phi_1 \left(k_y^2 - k_z^2\right) \sigma_x - \phi_2 \left(k_x^2 - k_z^2\right) \sigma_y$ \tabularnewline
$+ \, \phi_3 \left(k_x^2 - k_y^2\right) \sigma_z\bigg)$
\end{tabular} & \begin{tabular}{c}
$\lambda \left(\phi_1\left(\epsilon_{xy} H_y - \epsilon_{xz} H_z\right) - \phi_2\left(\epsilon_{xy} H_x - \epsilon_{yz} H_z\right) + \phi_3\left(\epsilon_{xz} H_x - \epsilon_{yz} H_y\right)\right)$ \\[20pt]
$\lambda' \left(\phi_1 \epsilon_{y^2-z^2} H_x - \phi_2 \epsilon_{x^2-z^2} H_y + \phi_3 \epsilon_{x^2-y^2} H_z\right)$
\end{tabular} \\[-12pt]
& & \tabularnewline

\hline
FM irrep & & \tabularnewline
\hline
${\color{FMirrepColour}T_{1g}^{-}}$ & \begin{tabular}{c}
\textcolor{FMirrepColour}{$g \left(\phi_1 \sigma_x + \phi_2 \sigma_y + \phi_3 \sigma_z\right)$}
\end{tabular} & \begin{tabular}{c}
\textcolor{FMirrepColour}{$\lambda \epsilon_{A_{1g}} \left(\phi_1 H_x + \phi_2 H_y + \phi_3 H_z\right)$}\tabularnewline
\textcolor{FMirrepColour}{$\lambda' \left(\phi_1\left(\epsilon_{xy} H_y + \epsilon_{xz} H_z\right) + \phi_2\left(\epsilon_{xy} H_x + \epsilon_{yz} H_z\right) + \phi_3\left(\epsilon_{xz} H_x + \epsilon_{yz} H_y\right)\right)$}\tabularnewline
\textcolor{FMirrepColour}{$\lambda'' \left(\phi_1 \epsilon_{y^2+z^2-2x^2} H_x + \phi_2 \epsilon_{x^2+z^2-2y^2} H_y + \phi_3\epsilon_{x^2+y^2-2z^2} H_z\right)$}\tabularnewline
\end{tabular}\tabularnewline
\hline \hline
\end{tabular}}
\caption{The coupling of altermagnetic (AM) and ferromagnetic (FM) order parameters $\phi$ to fermions and to simultaneous magnetic and strain fields (piezomagnetism) for the cubic point group $O_{h}$ ($m{-}3m$).
The first column indicates the irrep of $\phi$.
For $E_{g}$, $(\phi_1,\phi_2)$ transforms like $\big(x^2+y^2-2z^2,\sqrt{3}(x^2-y^2)\big)$, for $T_{2g}$ $(\phi_1,\phi_2,\phi_3)$ transforms like $(yz,zx,xy)$, while for $T_{1g}$ $(\phi_1,\phi_2,\phi_3)$ transforms like $(\sigma_x,\sigma_y,\sigma_z)$.
$g$ and $\lambda$ are coupling constants, $k_{\alpha}$ is the momentum, $\sigma_{\alpha}$ are Pauli matrices, $\epsilon_{\alpha\beta}$ is the strain tensor, and $H_{\alpha}$ is the magnetic field.
The possibility of having $g' \neq g$ and $\lambda' \neq \lambda$ reflects magnetic anisotropy.
$\epsilon_{A_{1g}}$ are the trivially transforming strain components.
The coupling to lattice fermions is deduced by replacing $k_{\alpha}$ with $\sin k_{\alpha}$.
Notice how the coupling for the altermagnetic order of $A_{1g}^{-}$, which forms a dotricontapolar or $32$-polar state, is possible only at high order in momentum and strains.}
\label{Tab3}
\end{table*}

\section{Critical elasticity for a cylindrical Fermi surface}
\label{App:CylindricalFS}

In the main text, we have assumed that the Fermi surface has the spherical shape. Here we consider the cylindrical Fermi surface characterised by $\xi_k=\frac{\mathbf{k}_{2d}^2}{2m}-\mu$. Due to the different Fermi surface geometry, the Landau damping for the boson is changed, while the one-loop correction from the elasticity is the same. For the dispersion of the boson, velocity along the $z$-direction is assumed to be much smaller than the velocity along the $xy$-plane. For the Sommerfeld coefficient of a cylindrical Fermi surface, the results are exactly the same as those for a spherical Fermi surface. Here, we include the calculation of the elastocaloric change, $\eta$. Since the elastocaloric change in the nematic case and the altermagnetic case with $\mathbf{H}=H\hat{z}$ are identical, we present the result for the nematic case only. However the altermagnet case with $\mathbf{H}=H\hat{x}$ is different from the nematic case so we include it.

\subsection{Landau damping}
Following the same procedure in Sec.~\ref{sec::alter_Landau},
\begin{align}
    \Delta D_i(q)&\equiv D_i(\Omega,\mathbf{q})-D_i(0,\mathbf{q})=\frac{2}{v_{\rm{F}}}\int\frac{dk_{2d}^2}{(2\pi)^2}\int \frac{dk_z}{2\pi}h_i^2(k)\delta(k_{2d}-k_{\rm{F}})\frac{1}{1-i\frac{v_{\rm{F}}\mathbf{q}_{2d}\cdot\hat{k}_{2d} }{\Omega}}.
\end{align}

Using the explicit expression of $\mathbf{h}(q)$, we obtain
\begin{align}
    \Delta D_x(q)&=\frac{2}{v_{\rm{F}}}\int \frac{d^2k_{2d}}{(2\pi)^2}\int_{-\pi/a_z}^{\pi/a_z}\frac{dk_z}{2\pi}\Big(\cos^2 k_x a-\cos^2 k_ya\Big)^2\frac{\delta(k_{2d}-k_{\rm{F}})}{1-iv_{\rm{F}}\mathbf{q}_{2d}\cdot \hat{k}_{2d}/\Omega}\nonumber\\
    &\approx \frac{(k_{\rm{F}}a)^4k_{\rm{F}}}{\pi v_{\rm{F}}a_z}\int\frac{d\theta_k}{2\pi}[\cos^2\theta_k-\sin^2\theta_k]^2\frac{1}{1-i\frac{v_{\rm{F}}|\mathbf{q}_{2d}|}{\Omega}\cos(\phi_q-\theta_k)}\nonumber\\
    &\approx  \frac{(k_{\rm{F}}a)^4k_{\rm{F}}}{2\pi v_{\rm{F}}a_z}\Big[(1+\cos4\phi_q)\frac{|\Omega|}{v_{\rm{F}} |\mathbf{q}_{2d}|}-4\cos4\phi_q \frac{\Omega^2}{v_{\rm{F}}^2|\mathbf{q}_{2d}|^2}\Big],
\end{align}
where $a$ and $a_z$ are lattice constants and $\mathbf{q}_{2d}=|\mathbf{q}_{2d}|(\cos\phi_q\hat{x}+\sin\phi_q\hat{x})$.

Similarly,
\begin{align}
    \Delta D_y(q)&\approx\frac{(k_{\rm{F}}a)^2k_{\rm{F}}}{4\pi v_{\rm{F}}a_z}\Big[(1-\cos2\phi_q)\frac{|\Omega|}{v_{\rm{F}}|\mathbf{q}_{2d}|}+2\cos 2\phi_q \frac{\Omega^2}{v_{\rm{F}}^2|\mathbf{q}_{2d}|^2}\Big],\\
    \Delta D_z(q)&\approx \frac{(k_{\rm{F}}a)^2k_{\rm{F}}}{4\pi v_{\rm{F}}a_z}\Big[(1+\cos2\phi_q)\frac{|\Omega|}{v_{\rm{F}}|\mathbf{q}_{2d}|}-2\cos 2\phi_q \frac{\Omega^2}{v_{\rm{F}}^2 |\mathbf{q}_{2d}|^2}\Big].
\end{align}

Note that the angle dependent terms of $\Delta D_y$ and $\Delta D_z$ have opposite signs and so they cancel each other.

Combining the above results, we can get
\begin{align}
    \Delta S_\phi^c\approx \int\frac{d^4q}{(2\pi)^4}\frac{1}{2}\phi(q)\Bigg[\frac{g_1^2k_{\rm{F}}(k_{\rm{F}}a)^4}{\pi v_{\rm{F}}a_z}\Bigg(\sin^22\theta_q \frac{|\Omega|}{v_{\rm{F}}|\mathbf{q}_{2d}|}+2\cos4\theta_q\frac{\Omega^2}{v_{\rm{F}}^2|\mathbf{q}_{2d}|^2}\Bigg)+\frac{g_2^2(k_{\rm{F}}a)^2k_{\rm{F}}}{2\pi v_{\rm{F}}a_z}\frac{|\Omega|}{v_{\rm{F}}|\mathbf{q}_{2d}|}\Bigg]\phi(-q),
\end{align}
where $\theta_q$ is the angle between $\mathbf{q}_{2d}$ and the unit vector $\hat{k}_1=\frac{\hat{x}+\hat{y}}{2}$ and $\theta_q=\phi_q-\frac{\pi}{4}$.

\subsection{Elastocaloric effect}
\subsubsection{Altermagnet case with $\mathbf{H}=H\hat{x}$}
For nematic systems and altermagnets with $H=H_z$, the results are the same as that of a spherical Fermi surface as the in-plane symmetry is unaffected, we therefore concentrate on $H=H_x$.
We use the following approximated mass $\bar{m}(\bar{q}_{2d},q_z,\epsilon,H)$ of $\mathbf{H}=H\hat{x}$ case
\begin{align}
    \bar{m}^2(\mathbf{q},\mathbf{H})\approx \left\{ \begin{array}{ll}
    \frac{\bar{q}_y^2+\bar{q}_z^2}{\bar{q}_x^2},& (\it{i})\;|q_x|\gg|q_y|,|q_z|\\
\frac{\bar{q}_x^2+\bar{q}_y^2}{\bar{q}_z^2}, & (\it{ii})\; |q_z|\gg |q_x|,|q_y|\\
1, & (\it{iii})\; |q_y|\gg |q_x|,|q_z|
\end{array}\right.\label{eq::nonZeroHxRenormalizedMassbar}
\end{align}

Using the spherical coordinates, setting the largest momentum to be $q\cos\theta$, we obtain following expressions for Eq.~\eqref{eq:f_efe_full}:
\begin{align}
    f^{(\it{i})}&\approx 2\int_0^\infty dx \int_0^\infty dq \int_0^{\theta_0}\int_0^{2\pi}d\phi \frac{\theta q^3\sqrt{1-\theta^2\cos^2\phi}x^2\mathrm{csch}^2(x/2)}{q^2[1-\theta^2\cos^2\phi]\Big(q^2[1-(1-c_z^2)\theta^2\cos^2\phi]+\theta^2\Big)^2+x^2\bar{T}^2},\\
    f^{(\it{ii})}&\approx 4\pi\int_0^\infty dx \int_0^{\theta_0}d\theta\int_0^\infty dq \frac{\theta^2q^3x^2\mathrm{csch}^2(x/2)}{q^2\theta^2\Big(q^2\theta^2+c_z^2q^2(1-\theta^2)+\theta^2\Big)^2+x^2\bar{T}^2},\\
    f^{(\it{iii})}&\approx 2\int_0^{2\pi} d\phi  \int_0^{\theta_0} d\theta \int_0^\infty dq\int_0^\infty dx\frac{x^2q^3\theta \sqrt{1-\theta^2\cos^2\phi}\mathrm{csch}^2(x/2)}{q^2(1-\theta^2\cos^2\phi)(q^2[1-(1-c_z^2)\theta^2\cos^2\phi]+1)^2+x^2\bar{T}^2}.
\end{align}

First let us consider the $f^{(\it{ii})}$. We divide the integral $\int_0^\infty dq$ into two parts and apply the approximation. It gives
\begin{align}
    f^{(\it{ii})}&=f_{[0,1]}^{(\it{ii})}+f_{[1,\infty]}^{(\it{ii})},\\
    f_{[0,1]}^{(\it{ii})}&\approx 4\pi\int_0^\infty dx \int_0^{\theta_0}d\theta\int_0^1 dq \frac{\theta^2q^3x^2\mathrm{csch}^2(x/2)}{q^2\theta^2[\theta^2+c_z^2q^2]^2+x^2\bar{T}^2},\\
    f_{[1,\infty]}^{(\it{ii})}&\approx 4\pi\int_0^\infty dx \int_0^{\theta_0}d\theta\int_1^\infty dq \frac{\theta^2q^3x^2\mathrm{csch}^2(x/2)}{c_z^4q^6\theta^2+x^2\bar{T}^2}.
\end{align}

At high temperature $\bar{T}\gg1$, the $f^{(\it{ii})}_{[1,\infty]}$ is dominant term while the $f^{(\it{ii})}_{[0,1]}$ is dominant one in low temperature $\bar{T}\ll 1$. It can be seen clearly by rescale the $q$ and $\theta$ to eliminate $T$ dependence in the denominator. 

At $T=0$, the term $f^{(\it{ii})}_{[0,1]}$ diverges. For the finite temperature, both $f^{(\it{ii})}_{[0,1]}$ and $f^{(\it{ii})}_{[1,\infty]}$ show the $T^{-2/3}$ behaviour. As a result, 
\begin{align}
    f^{(\it{ii})}\approx \bar{T}^{-2/3}.
\end{align}

To evaluate the $f^{(\it{i})}$ and $f^{(\it iii)}$, we first use Taylor expansion with respect to $\theta^2\cos^2\phi$ and consider only zeroth order terms (Higher order terms give sub-leading contributions). Then,
\begin{align}
    f^{(\it{i})}&\approx 2\int_0^\infty dx \int_0^\infty dq \int_0^{\theta_0}\int_0^{2\pi}d\phi \frac{\theta q^3x^2\mathrm{csch}^2(x/2)}{q^2\Big(q^2+\theta^2\Big)^2+x^2\bar{T}^2},\\
    f^{(\it{iii})}&\approx 2\int_0^{2\pi}d\phi \int_0^{\theta_0}d\theta \int_0^\infty dx \int_0^\infty dq \frac{x^2q^3\theta\mathrm{csch}^2(x/2)}{q^2[q^2+1]^2+x^2\bar{T}^2}.
\end{align}

Now repeating the same procedure done for the evaluation of $f^{(\it ii)}$, we can get following qualitative behaviour of the $f^{(\it i)}$ and $f^{(\it{iii})}$ in the limit temperatures as follows:
\begin{align}
f^{(\it{i})}&\approx  \begin{cases}-\log\bar{T} & \bar{T}\ll 1 \\
 \bar{T}^{-2/3} & \bar{T}\gg 1\end{cases},\qquad f^{(\it{iii})}\approx  \begin{cases}\rm{const}, & \bar{T}\ll 1 \\
 \bar{T}^{-2/3}, & \bar{T}\gg 1\end{cases}
\end{align}

Combining the contributions from $(\it i)$, $(\it ii)$ and $(\it iii)$ gives the following elastocaloric change:
\begin{align}
 \eta\approx - 144\pi b \Gamma_{*,0}T_{*} \frac{\left(T/T_{*}\right)^{1/3}}{\log\frac{T_{0}}{T}}
\end{align}
 Unlike to the spherical Fermi surface case, $\eta$ in the cylindrical Fermi surface with $\mathbf{H}=H\hat{x}$ shows the $\frac{T^{1/3}}{\log T}$ behaviour not only at the high temperature regime but also at the low energy regime.

\end{widetext}

\end{document}